\documentclass{article}

\usepackage{tikz}
\usepackage{amsmath}

\usepackage{filecontents}

\usepackage{tikz}
\usepackage{amsmath} 

\usepackage{paralist}

\usepackage{subcaption}
\usepackage{graphicx} 

\usepackage{array}
\usepackage{ragged2e}
\newcolumntype{P}[1]{>{\RaggedRight\hspace{0pt}}p{#1}}

\usepackage{amsmath}
\usepackage{amsthm}
\usepackage{url}
\usepackage{comment}
\usepackage{listings}
\usepackage[framemethod=tikz]{mdframed}
\usepackage{multirow}
\usepackage{pdflscape}
\usepackage{soul}

\newcommand{\Bank}{\ensuremath{B}}

\newcommand{\datafont}[1]{\text{#1}}
\newcommand{\cmdfont}[1]{\ensuremath{\mathsf{#1}}}
\newcommand{\hexfont}[1]{\ensuremath{\mathtt{#1}}}

\newcommand{\PDOL}{\datafont{PDOL}}

\newcommand{\expdate}{\datafont{expDate}}

\newcommand{\CVM}{\datafont{CVM}}

\newcommand{\AIP}{\datafont{AIP}}

\newcommand{\AFL}{\datafont{AFL}}
\newcommand{\ATC}{\datafont{ATC}}

\newcommand{\CDOL}[1]{\datafont{CDOL#1}}

\newcommand{\CID}{\datafont{CID}}
\newcommand{\AID}{\datafont{AID}}
\newcommand{\IAD}{\datafont{IAD}}

\newcommand{\AC}{\datafont{AC}}
\newcommand{\SDAD}{\datafont{SDAD}}

\newcommand{\PAN}{\datafont{PAN}}

\newcommand{\CTQ}{\datafont{CTQ}}

\newcommand{\Ks}{\ensuremath{s}}

\newcommand{\MAC}[2]{\mathit{MAC}_{#2}(#1) }

\definecolor{codegreen}{rgb}{0,0.6,0}
\definecolor{codegray}{rgb}{0.5,0.5,0.5}
\definecolor{codepurple}{rgb}{0.58,0,0.82}
\definecolor{backcolour}{rgb}{0.98,0.98,0.95}
\definecolor{codebrown}{rgb}{0.5,0.0,0.0}

\newcommand{\nicetilde}{\raisebox{0.5ex}{\texttildelow}}

\lstdefinelanguage{tamarin}{
  morekeywords={lemma, All, Ex, not, exists, trace, rule, In, Out, let},
  sensitive=false,
  morecomment=[l]{//},
  morestring=[b]',
  literate={@}{{{\color{codebrown}@{}}}}1%
           {=}{{{\color{codebrown}={}}}}1%
           {<}{{{\color{codebrown}<{}}}}1%
           {>}{{{\color{codebrown}>{}}}}1%
           {[}{{{\color{codebrown}[{}}}}1%
           {]}{{{\color{codebrown}]{}}}}1%
           {|}{{{\color{codebrown}|{}}}}1%
           {\#}{{{\color{codebrown}\#{}}}}1%
           {:}{{{\color{codebrown}:{}}}}1%
           {\&}{{{\color{codebrown}\&{}}}}1%
           {-}{{{\color{codebrown}-{}}}}1
           {~}{{{\nicetilde{}}}}1
}

\lstdefinelanguage{tlog}{
  morecomment=[l]{//}
}

\newcommand{\z}[1]{\mathsf{#1}}

\usepackage[acronym, nomain, nohypertypes={acronym}]{glossaries}

\makeglossaries

\newacronym{AAC}{AAC}{Application Authentication Cryptogram}
\newacronym{AC}{AC}{Application Cryptogram}
\newacronym{ADF}{ADF}{Application Definition File}
\newacronym{AFL}{AFL}{Application File Locator}
\newacronym{AID}{AID}{Application Identifier}
\newacronym{AIDs}{AIDs}{Application Identifiers}
\newacronym{AIP}{AIP}{Application Interchange Profile}
\newacronym{APDU}{APDU}{Application Protocol Data Unit}
\newacronym{ARC}{ARC}{Authorization Response Code}
\newacronym{ARPC}{ARPC}{Authorization Response Cryptogram}
\newacronym{ARQC}{ARQC}{Authorization Request Cryptogram}
\newacronym{ATC}{ATC}{Application Transaction Counter}
\newacronym{AUC}{AUC}{Application Usage Control}

\newacronym{BCD}{BCD}{Binary Coded Decimal}
\newacronym{BDH}{BDH}{Blinded Diffie-Hellman}

\newacronym{CDA}{CDA}{Combined DDA/Application Cryptogram}
\newacronym{CDOL}{CDOL}{Card Risk Management Data Object List}
\newacronym{CID}{CID}{Cryptogram Information Data}
\newacronym{CPR}{CPR}{Card Processing Requirements}
\newacronym{CTQ}{CTQ}{Card Transaction Qualifiers}
\newacronym{CVC3}{CVC3}{Dynamic Card Verification Code}
\newacronym{CVM}{CVM}{Cardholder Verification Method}
\newacronym{CVMR}{CVMR}{Cardholder Verification Method Results}
\newacronym{CVR}{CVR}{Cardholder Verification Rules}
\newacronym{CDCVM}{CDCVM}{Consumer Device CVM}
\newacronym{CSC}{CSC}{Card Security Code}
\newacronym{CVV}{CVV}{Card Verification Value}
\newacronym{CVC}{CVC}{Card Verification Code}
\newacronym{CVD}{CVD}{Cardholder Verification Decision}
\newacronym{CMC}{CMC}{Card Message Counter}
\newacronym{CA}{CA}{Certificate Authority}

\newacronym{DDA}{DDA}{Dynamic Data Authentication}
\newacronym{DDF}{DDF}{Dictionary Definition File}
\newacronym{DOL}{DOL}{Data Object List}
\newacronym{DDOL}{DDOL}{Dynamic Data Object List}
\newacronym{DH}{DH}{Diffie–Hellman}
\newacronym{DRRE}{DRRE}{Device Relay Resistance Entropy}
\newacronym{DoS}{DoS}{Denial of Service}

\newacronym{ECC}{ECC}{Elliptic Curve Cryptography}
\newacronym{EDA}{EDA}{Enhanced Data Authentication}

\newacronym{fDDA}{fDDA}{fast Dynamic Data Authentication}
\newacronym{FCI}{FCI}{File Control Information}

\newacronym{GPO}{GPO}{Get Processing Options}

\newacronym{HCE}{HCE}{Host-based Card Emulation}

\newacronym{IAC}{IAC}{Issuer Action Code}
\newacronym{IAD}{IAD}{Issuer Application Data}
\newacronym{ICC}{ICC}{Integrated Circuit Card}

\newacronym{KDF}{KDF}{Key Derivation Function}

\newacronym{MCC}{MCC}{Merchant Category Code}
\newacronym{MITM}{MITM}{machine-in-the-middle}
\newacronym{MAC}{MAC}{Message Authentication Code}

\newacronym{NFC}{NFC}{Near Field Communication}

\newacronym{ODA}{ODA}{Offline Data Authentication}
\newacronym{OTA}{OTA}{Online Transaction Authorization}

\newacronym{PAN}{PAN}{Primary Account Number}
\newacronym{PDOL}{PDOL}{Processing Data Object List}
\newacronym{PKI}{PKI}{Public Key Infrastructure}
\newacronym{PoS}{POS}{Point-of-Sales}
\newacronym{POS}{POS}{Point-of-Sales}
\newacronym{PSE}{PSE}{Payment System Environment}
\newacronym{PPSE}{PPSE}{Proximity Payment System Environment}
\newacronym{PIN}{PIN}{Personal Identification Number}
\newacronym{PK}{PK}{Public Key}

\newacronym{qVSDC}{qVSDC}{quick Visa Smart Debit/Credit}

\newacronym{RID}{RID}{Registered Application Provider Identifier}
\newacronym{RRP}{RRP}{Relay Resistance Protocol}

\newacronym{SDA}{SDA}{Static Data Authentication}
\newacronym{SDAD}{SDAD}{Signed Dynamic Authentication Data}
\newacronym{SDR}{SDR}{Software Defined Radio}
\newacronym{SFI}{SFI}{Short File Identifier}
\newacronym{SSAD}{SSAD}{Signed Static Authentication Data}
\newacronym{SSL/TLS}{SSL/TLS}{Secure Sockets Layer/Transport Layer Security}
\newacronym{SK}{SK}{Secret Key}

\newacronym{TA}{TA}{Transaction Authorization}
\newacronym{TAC}{TAC}{Terminal Action Code}
\newacronym{TC}{TC}{Transaction Cryptogram}
\newacronym{TDHC}{TDHC}{Transaction Data Hash Code}
\newacronym{TT}{TT}{Terminal Type}
\newacronym{TTQ}{TTQ}{Terminal Transaction Qualifiers}
\newacronym{TVR}{TVR}{Terminal Verification Results}
\newacronym{TRRE}{TRRE}{Terminal Relay Resistance Entropy}

\newacronym{UN}{UN}{Unpredictable Number}
\newacronym{UML}{UML}{Unified Modeling Language}
\newacronym{UID}{UID}{Unique Identifier}

\newacronym{VSDC}{VSDC}{Visa Smart Debit/Credit}

\usepackage{msc}

\newlength{\xtmpwidth}
\newlength{\xtmpheight}

\setmscvalues{large} %
\setmsckeyword{}
\drawframe{no}

\newlength{\var}

\newcommand{\boxtext}[1]{
\lower 3pt \hbox{%
  \vbox{%
    \hrule
    \hbox{\strut \vrule{} #1 \vrule}%
    \hrule
  }%
  }%
}

\newlength{\leftprotobox}
\newlength{\middleprotobox}
\newlength{\rightprotobox}
\newlength{\protoboxheight}
\newcommand{\toleft}[3]{
\par\smallskip
\centerline{
\hbox{%
    \vbox{\vfil\hbox to \leftprotobox{\hfil #1\hfil}\vfil}
    \vbox{\hbox to \middleprotobox{\hfil #2\hfil} \vskip -\protoboxheight \hbox to \middleprotobox{:=fill}}
    \vbox{\vfil\hbox to \rightprotobox{\hfil #3 \hfil}\vfil}
   }%
  }
\smallskip
}

\newcommand{\toright}[3]{
\par\smallskip
\centerline{
\hbox{%
    \vbox{\vfil\hbox to \leftprotobox{\hfil #1\hfil}\vfil}
    \vbox{\hbox to \middleprotobox{\hfil #2\hfil} \vskip -\protoboxheight \hbox to \middleprotobox{\rightarrowfill}}
    \vbox{\vfil\hbox to \rightprotobox{\hfil #3\hfil}\vfil}
    }%
   }
\smallskip
}

\newcommand{\noarrow}[3]{
\par\smallskip
\centerline{
\hbox{%
    \vbox{\hbox to \leftprotobox{\hfil #1\hfil}}
    \vbox{\hbox to \middleprotobox{\hfil#2\hfil}}
    \vbox{\hbox to \rightprotobox{\hfil #3 \hfil}}
           }%
}
\smallskip
}

{%

}

\newcommand{\dataf}[1]{\textbf{#1}}

\newcommand{\ACx}{\dataf{AC}}
\newcommand{\ACfull}{\acrlong{AC}~(\dataf{AC})}
\newcommand{\AIDx}{\dataf{AID}}
\newcommand{\AIDxpl}{\dataf{AID}s}
\newcommand{\AIDfull}{\acrlong{AID}~(\dataf{AID})}

\newcommand{\PDOLfull}{\acrlong{PDOL}~(\dataf{PDOL})}
\newcommand{\PDOLdatax}{\dataf{PDOL Related Data}}
\newcommand{\UNTx}{\dataf{UN$_\text{T}$}}

\newcommand{\UNCx}{\dataf{UN$_\text{C}$}}

\newcommand{\AIPx}{\dataf{AIP}}
\newcommand{\AIPfull}{\acrlong{AIP}~(\dataf{AIP})}
\newcommand{\AFLx}{\dataf{AFL}}
\newcommand{\AFLfull}{\acrlong{AFL}~(\dataf{AFL})}
\newcommand{\PANx}{\dataf{PAN}}

\newcommand{\CDOLx}{\dataf{CDOL}}
\newcommand{\CDOLfull}{\acrlong{CDOL}~(\dataf{CDOL})}

\newcommand{\CDOLOneDatax}{\dataf{CDOL1 Related Data}}
\newcommand{\CDOLDatax}{\dataf{CDOL Related Data}}
\newcommand{\CVMListx}{\dataf{CVM List}}
\newcommand{\mkx}{\dataf{mk}}

\newcommand{\ATCx}{\dataf{ATC}}
\newcommand{\ATCxpl}{\dataf{ATC}s}
\newcommand{\ATCfull}{\acrlong{ATC}~(\dataf{ATC})}
\newcommand{\SDADx}{\dataf{SDAD}}
\newcommand{\SDADfull}{\acrlong{SDAD}~(\dataf{SDAD})}

\newcommand{\IADx}{\dataf{IAD}} 
\newcommand{\IADfull}{\acrlong{IAD}~(\dataf{IAD})}

\newcommand{\CVMResultsx}{\dataf{CVM Results}}
\newcommand{\TTQx}{\dataf{TTQ}}
\newcommand{\TTQfull}{\acrlong{TTQ}~(\dataf{TTQ})}
\newcommand{\CTQx}{\dataf{CTQ}}
\newcommand{\CTQfull}{\acrlong{CTQ}~(\dataf{CTQ})}

\newcommand{\TrackTwodatax}{\dataf{Track~2 Data}}
\newcommand{\TrackOneTwodatax}{\dataf{Track~1} and \dataf{Track~2 Data}}
\newcommand{\TrackTwoeqdatax}{\dataf{Track~2 Equivalent Data}}
\newcommand{\TrackOneTwoeqdatax}{\dataf{Track~1} and \dataf{Track~2 Equivalent Data}}

\newcommand{\CVCthreex}{\dataf{CVC3}}
\newcommand{\CVCthreexpl}{\dataf{CVC3}s}
\newcommand{\CVCthreefull}{\acrlong{CVC3}~(\dataf{CVC3})}
\newcommand{\CSCx}{\dataf{CSC}}
\newcommand{\CSCfull}{\acrlong{CSC}~(\dataf{CSC})}
\newcommand{\TACDenialx}{\dataf{TAC-Denial}}
\newcommand{\TACDenialfull}{\acrlong{TAC}-Denial (\dataf{TAC-Denial})}
\newcommand{\TACOnlinex}{\dataf{TAC-Online}}

\newcommand{\TACDefaultx}{\dataf{TAC-Default}}

\newcommand{\IACdenialx}{\dataf{IAC-Denial}}
\newcommand{\CApkIndexx}{\dataf{CA Public Key Index}}
\newcommand{\CApkx}{\dataf{CA Public Key}}
\newcommand{\TVRx}{\dataf{TVR}}
\newcommand{\TVRfull}{\acrlong{TVR}~(\dataf{TVR})}
\newcommand{\MCCx}{\dataf{MCC}}
\newcommand{\MCCfull}{\acrlong{MCC}~(\dataf{MCC})}

\usepackage{enumitem}
\usepackage{multicol}

\usepackage{hyperref}
\usepackage[hyphenbreaks]{breakurl}

\newcommand{\secref}[1]{\S\ref{#1}}
\newcommand{\secreftwo}[2]{\S\ref{#1} and~\S\ref{#2}}
\newcommand{\secrefthree}[3]{\S\ref{#1}, \S\ref{#2}, and \S\ref{#3}}
\newcommand{\secreffour}[4]{\S\ref{#1}, \S\ref{#2}, \S\ref{#3}, and \S\ref{#4}}

\begin{document}

\date{Version 1.0\\April 4, 2025}

\title{\Large \bf SoK: Attacks on Modern Card Payments}

\author{
  {\rm Xenia Hofmeier$^1$, David Basin$^1$, Ralf Sasse$^1$, and Jorge Toro-Pozo$^2$}\\
  $^1$ Department of Computer Science, ETH Zurich, Switzerland \\
  \{xenia.hofmeier, basin, ralf.sasse\}@inf.ethz.ch \\
  $^2$ SIX Digital Exchange, Switzerland\\
  jorge.toro@sdx.com
  }

\maketitle

\begin{abstract}
EMV is the global standard for smart card payments
and its NFC-based version, EMV contactless, is widely used, also for mobile payments.
In this systematization of knowledge, we examine attacks on the EMV contactless protocol.
We provide a comprehensive framework encompassing its desired security properties and adversary models.
We also identify and categorize a comprehensive collection of protocol
flaws and show how different subsets thereof can be combined into attacks.
In addition to this systematization, we examine the underlying reasons
for the many attacks against EMV and point to a better way forward.
\end{abstract}

\section{Introduction}\label{Section:Introduction}

EMV, named after its founding organizations Europay, Mastercard, and Visa, is the de facto protocol standard for smart card payments.
The protocol specifications are 
maintained by EMVCo, a consortium consisting of American Express, Discover, JCB, Mastercard, UnionPay, and Visa~\cite{OrganisationStructure}.
With 12.8 billion EMV cards, EMV transactions constitute 94\% of card-present chip transactions~\cite{WhyEMVa}.
Its central role in worldwide payments makes EMV highly security critical.

EMV consists of a family of payment technologies, and the most well known are EMV contact and contactless.
In a contact transaction, the card is inserted into 
a payment terminal, whereas contactless cards communicate with terminals wirelessly over NFC.
Mobile phone-based card payments, like Apple Pay, Samsung Pay, and Google Pay, also use this protocol over the phone's NFC interface.
The contactless protocol is based on the older contact protocol and there are eight so-called kernels, which are variants of the contactless protocol,
each associated with a different EMVCo member.
The protocol is highly complex given the eight kernels, numerous configuration options, and a specification of over 2500 pages, spanning 15 books with many cross-references.

\looseness=-1
The current chip-based payment cards' predecessors, magnetic stripe cards, 
used just static data for payments, making them susceptible to simple attacks such as cloning attacks. 
The integrated circuits of EMV contact cards prevent such attacks 
with dynamically generated data in 
each transaction.
However, EMV contact is still vulnerable to attacks, the most 
prominent being the \acrfull{MITM} attack 
that enables the adversary to pay
with a card while providing the wrong PIN~\cite{murdochChipPINBroken2010}.
However, 
the attack's bulky, wired \acrshort{MITM} infrastructure
is difficult to conceal and makes the attack impractical.
In contrast, 
the various attacks on EMV contactless target
the contactless cards' wireless interface, which can be easily and discretely
accessed by the attacker.
The introduction of phone-based payments also allows for an inconspicuous implementation of a card emulator where,  for store employees, a normal phone payment is indistinguishable from 
a malicious emulation
on a smartphone.
This makes attacks on EMV contactless both inconspicuous and easy to carry out. 

We focus in this SoK on modern payment cards.
Thus, we focus on chip-based payment cards rather than older technologies
like magnetic stripes.
Moreover, from all the chip-based standards, we focus on EMV, as it specifies the most prevalent 
chip transactions.
We further restrict our scope to EMV contactless as it is the newest standard, has a large attack surface, and is still evolving.
Moreover, the payment industry's push for 
EMV contactless has led to its wide adoption.
Thus, attacks on EMV contactless are highly
relevant.

Note that payment cards are used in a wide range of transaction types.
Most payment cards have interfaces for contactless and contact EMV as well as magnetic stripe transactions.
The information printed on the card can be used for card-not-present transactions, for example over the Internet or phone.
And cards can be used at ATMs or as a second factor for e-banking.
In this SoK, we focus on attacks targeting EMV contactless transactions, and we therefore do not consider EMV technologies~\cite{EMVTechnologies} such as EMV contact, 3-D secure, and Payment Tokenisation.
We also focus on attacks targeting protocol weaknesses.
Thus, we do not consider social engineering attacks, ATM skimmers, or weaknesses in the bank's processes, such as weak second factor authentication or problematic card lock policies as presented in~\cite{anwarWalletWeTrust2024}.

\paragraph{\textbf{Contribution}}

\looseness=-1
In this SoK, we summarize and categorize attacks on EMV contactless.
We describe this protocol, its security requirements, relevant adversary models, and its flaws.

We provide a comprehensive framework to evaluate the security of EMV contactless consisting of security properties and 
adversary models.
This is important to determine the practicality and
relevance of attacks, 
i.e., whether they 
violate 
relevant properties and 
depend on
realistic adversaries. 

By extensively categorizing flaws and attacks on EMV contactless, we gain an understanding of the 
flaws discovered and
how they are composed into different attacks. 
This comprehensive list 
helps identifying 
new flaws and attacks,
understanding how they arise, and how we can improve the standard going forward.

\looseness=-1  
By studying past flaws and exploits, we also gain an understanding of their underlying causes. Examples include the trade-off between preventing attacks and having high acceptance rates,
and the need for backward compatibility.
We compare the approaches used to find attacks and explain the importance of both formal analysis and
testing on live systems.

\paragraph{\textbf{Outline}}
In \secref{Section:Background}, we provide background on EMV contactless.
In \secref{Section:SecurityProperties}, we identify security properties that are desirable for EMV contactless.
In \secref{Section:AdversaryCapabilities}, we present a family of adversary models and explain how the adversary capabilities are used in practice.
In \secref{Section:Flaws}, we categorize the flaws that enable the attacks  described in \secref{Section:AttacksOnEMV}.
In \secref{Section:Discussion}, we discuss the identified flaws, their possible causes, and the attack discovery process, and make suggestions for the future.
In \secref{Section:Conclusion}, we draw conclusions.

\section{EMV Contactless Protocol}\label{Section:Background}

\looseness=-1
In this section, we summarize the EMV contactless protocol~\cite{emvcoEMVContactlessSpecifications2023i}.
We simplify some aspects, focusing on the features that are relevant for the attacks discussed in this SoK.
We first describe a general EMV contactless transaction.
Afterwards we highlight some features of different versions: the Mastercard kernel, Visa kernel, and transactions with mobile devices.

\subsection{Protocol Overview}\label{Section:ProtocolOverview}

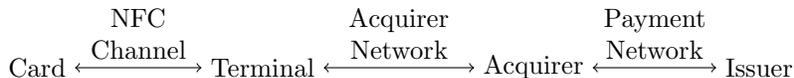
\begin{figure}[tb]
    \centering
        \begin{tikzpicture}

    \node (Card) at (-4.8,0.3) {Card};
    \node (Terminal) at (-1.8,0.3) {Terminal};
    \node (Acquirer) at (1.8,0.3) {Acquirer};
    \node (Issuer) at (4.8,0.3) {Issuer};

    \draw[<->] (Card.east) -- node[above] {\parbox{2cm}{\centering NFC\\ Channel}} (Terminal.west);
    \draw[<->] (Terminal.east) -- node[above] {\parbox{2cm}{\centering Acquirer\\ Network}} (Acquirer.west);
    \draw[<->] (Acquirer.east) -- node[above] {\parbox{2cm}{\centering Payment\\ Network}} (Issuer.west);

\end{tikzpicture}
    \caption{System Model.}
    \label{figure:SysModel}
\end{figure}

An EMV transaction allows a cardholder to pay merchants. 
To do this, the cardholder presents their card 
to the merchant's payment terminal.\footnote{We use the term ``terminal'' for the entire \acrfull{POS} System, consisting of the reader and terminal, as described in~\cite{emvcoEMVContactlessSpecifications2023i}.}
Figure~\ref{figure:SysModel} depicts the parties involved in an EMV contactless transaction and their communication channels.
The protocol specifies 
which messages are sent 
between the \emph{card} and the \emph{terminal} and 
between the terminal and the financial institution that issued the card, called the issuing bank or just \emph{issuer}. 
Two additional parties are involved in the communication between the terminal and the issuer: 
the financial institution that processes card payments on behalf of the merchant, called the acquiring bank or just \emph{acquirer}, and the \emph{payment network}, e.g. Visa or Mastercard, which connects the acquirer and issuer~\cite{basinCardBrandMixup2021}. 

\looseness=-1
The EMV contactless protocol serves two main purposes.
First, the issuer should be convinced that the cardholder agreed to the transaction so that the issuer can transfer the funds to the acquirer, and ultimately the merchant.
Second, the merchant should be convinced that the presented card is legitimate and that the transaction was successful so that the merchant can accept the payment.
These two purposes are achieved using two different cryptographic mechanisms.
The card convinces the issuer using a MAC.
This MAC is based on a symmetric key shared between the card and the issuer that is set up during the 
issuing process.
The merchant is convinced by a signature, 
produced with the card's private key.
The card's public key is authenticated using
a global \acrfull{PKI}.
We call the card's authentication to the issuer \acrfull{OTA}, as it requires the terminal to go \emph{online} to send the MAC to the issuer, and the authentication to the terminal \acrfull{ODA}, as the terminal can verify the signature \emph{offline}.

\looseness=-1
After verifying the signature, the terminal can either decline the transaction, accept it offline or send the transaction including the MAC to the issuer for online authorization.
In case of offline acceptance, 
the transaction data and MAC are sent later to the issuer.
Thus, the terminal accepts a transaction before the issuer could verify the MAC.
So, by accepting the transaction offline, the merchant risks that the transaction is later declined by the issuer, see~\secref{SubSec:Attacks:MerchantHoldingBag}.
Offline accepted transactions can be useful in situations with limited internet access.

EMV contactless consists of eight protocol variants, called kernels. 
Kernels~2~-~7 are for the six members of EMVCo and kernel 1 is no longer in use.
The new kernel~8 is intended to unify all existing kernels in the future.

There are two EMV contactless modes: EMV mode and mag-stripe mode.
EMV mode is similar to EMV contact and mag-stripe mode transmits data 
similar to 
the data on magnetic stripes. 
However, mag-stripe mode is not to be confused with the physical magnetic stripes on payment cards.
Mag-stripe mode's data merely resembles the data on the magnetic stripe.
It is a mode of EMV contactless and thus operates over NFC.
As EMV mode is 
more modern,
we focus here on EMV mode and present
mag-stripe mode in Appendix~\secref{SubSec:Protocol:MagStripeMode}.

\subsection{Protocol Phases}\label{Section:DetailsOnEMV}\label{SubSec:ProtocolPhases}

\begin{figure}[tb] 
    \centering
    
    \input{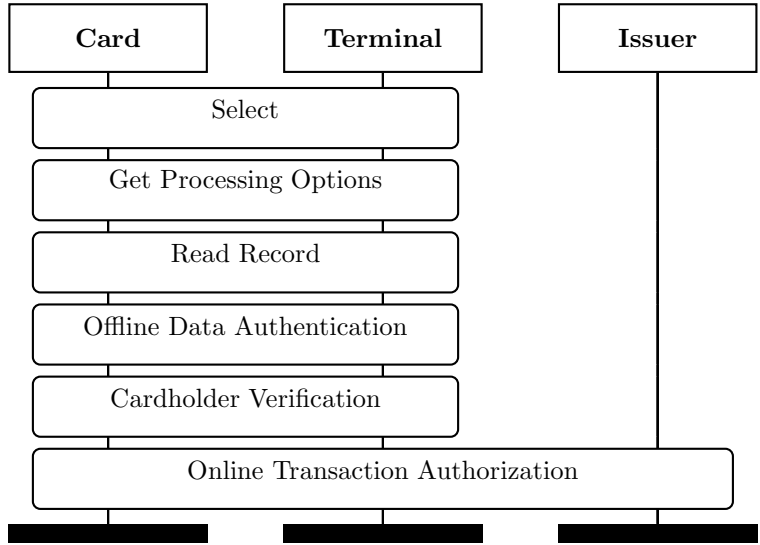}
    
    \caption{Phases of an EMV contactless transaction. }\label{figure:overviewEMV}
\end{figure}

We split the protocol into six phases, presented in Figure~\ref{figure:overviewEMV}. 
These phases are usually
executed in this sequence but might overlap in practice.
In the Select phase, the card and the terminal agree on a kernel 
that they subsequently run.
In the Get Processing Options phase, the terminal sends transaction-specific data 
and the card sends its capabilities.
In the Read Record phase, the card sends its static data.
The card authenticates to the terminal with \acrfull{ODA},
and the card authenticates 
to the issuer with \acrfull{OTA}. 
The cardholder's identity is verified using a \acrfull{CVM}, such as a \acrshort{PIN}.
We now provide more details on the protocol phases.
Each phase consists of terminal commands followed by card responses.
We highlight data using \dataf{bold font}.\footnote{Unfortunately the EMV standard contains many acronyms.  We have tried to minimize their usage, but stick to the essential ones to be faithful to the standard and to set the scene for analyzing attacks.}

\begin{figure*}[tb]  
    \centering
    \input{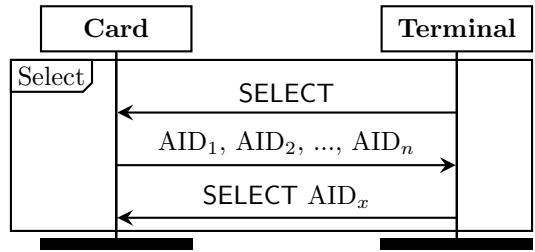}
    \caption{Select phase.}
    \label{figure:Protocol:msc:Select}
\end{figure*}

\subsubsection{Select Phase \textmd{(Figure~\ref{figure:Protocol:msc:Select})}}

The terminal and card agree on an application to run, called ``Combination'' in~\cite{emvcoEMVContactlessSpecifications2023i}.
The application identifies the kernel 
to be run and the configuration data 
to be used. 
The terminal requests the applications that the card supports
by sending a SELECT command.
The card answers with a list of supported applications.
The terminal chooses one of the applications
and replies with a SELECT \AIDx{} command,
where the \AIDfull{} identifies the chosen application.

\begin{figure*}[tb] 
    \centering 
    \input{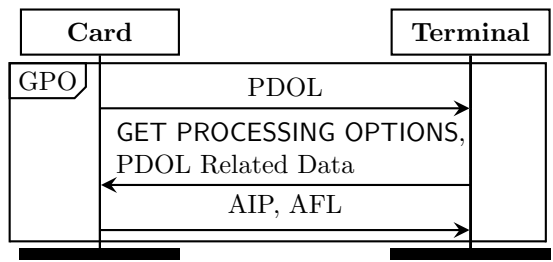}
    \caption{Get Processing Options phase}
    \label{figure:Protocol:msc:GPO}
\end{figure*}

\subsubsection{Get Processing Options \textmd{(Figure~\ref{figure:Protocol:msc:GPO})}} 

The card replies to the SELECT command with the \PDOLfull{}.
In general, 
\acrfullpl{DOL} 
specify the data elements that the card requests. 
The terminal provides this requested data, the \PDOLdatax{}, with the GET PROCESSING OPTIONS command. 
It contains transaction data such as the 
amount, 
and a terminal-generated nonce \UNTx{}. 
The card's reply contains the \AIPfull{} and \AFLfull{}. 
The \AIPx{} 
indicates
the card's capabilities.
The \AFLx{} identifies the files
that contain 
the card's static data.

\begin{figure*}[tb]  
    \centering
    \input{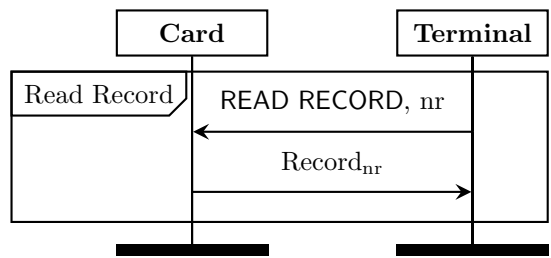}
    \caption{Read Record phase}
    \label{figure:Protocol:msc:ReadRecord}
\end{figure*}

\subsubsection{Read Record \textmd{(Figure~\ref{figure:Protocol:msc:ReadRecord})}} 

The terminal requests the card's static data, 
identified by the \AFLx{}.
The terminal sends a READ RECORD command with a record number and the card responds with the requested record. 
The terminal repeats this 
for each record in the \AFLx{}. 
The terminal usually requests the card's \PANx{} (i.e. the card number), expiration date, and optionally the card's public key and certificates, the \CDOLfull{}, and the \CVMListx. 
With the \CDOLx{}, the card requests the \CDOLDatax{}, which contains transaction data, and the \CVMListx{} indicates the card's supported CVMs.

\subsubsection{Offline Data Authentication}\label{SubSec:Protocol:ODA}

The card authenticates to the terminal in the \acrfull{ODA} phase using the card-generated signature \emph{\SDADfull{}}.
Performing \acrshort{ODA} is not required for all transactions and by all kernels, and some might require different \acrshort{ODA} methods.

To create the signature, the card has an asymmetric key pair.
The public key is authenticated using a global \acrshort{PKI}, where root certificate authorities provide certificates to the issuers, who create certificates for their issued cards. 
The card sends during the Read Record phase its certificate, the issuer's certificate, and a pointer to the root certificate, called the \CApkIndexx{}.
The terminal uses this pointer to look up 
the root \acrshort{CA}'s certificate in its database to verify the card's signature
and certificates. 
Note that 
the terminal relies solely on the 
\CApkIndexx{} to look up the \CApkx{}.
If the card offers an invalid index, the terminal fails to select a root CA.

The most basic \acrshort{ODA} method is \acrfull{SDA}, which relies on the issuer's signature on static data.
Alternatively, \acrfull{DDA} relies on the card's signature on dynamic data,
such as the 
terminal- and card-sourced nonces \UNTx{} and \UNCx{}.
The \acrshort{DDA}'s \SDADx{} does not authenticate transaction details.
In contrast, \acrfull{CDA} also authenticates transaction details.
We now highlight the \acrshort{ODA} methods used by Visa and Mastercard, namely \acrshort{fDDA} and \acrshort{CDA}.

\begin{figure*}[tb]  
    \centering
    \input{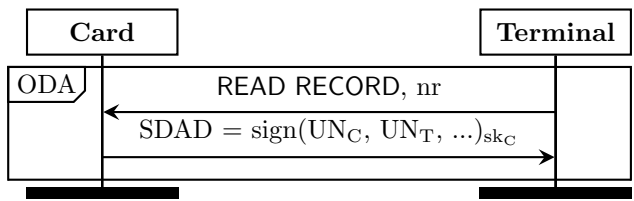}
    \caption{Overview of the \acrshort{ODA} method \acrshort{fDDA}. }
    \label{figure:Protocol:msc:ODA:fDDA}
\end{figure*}

\textbf{Visa's \acrshort{fDDA}} (Figure~\ref{figure:Protocol:msc:ODA:fDDA}). 
Visa's \acrfull{fDDA} is based on \acrshort{DDA} and supports high transaction speeds.
\acrshort{fDDA}'s \SDADx{}
contains the card's and terminal's nonces \UNCx{} and \UNTx{} as well as transaction and control data.
The card sends the \SDADx{} as READ RECORD response.

\begin{figure*}[tb]  
    \centering
    \input{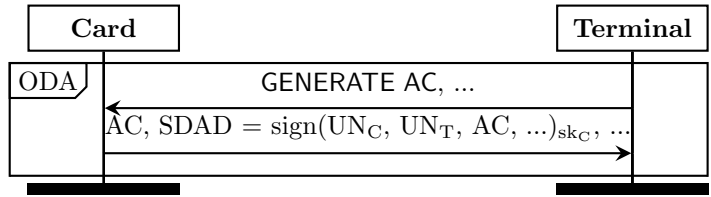}
    \caption{Overview of the \acrshort{ODA} method \acrshort{CDA}.  }
    \label{figure:Protocol:msc:ODA:CDA}
\end{figure*}

\textbf{Mastercard's \acrshort{CDA}} (Figure~\ref{figure:Protocol:msc:ODA:CDA}).
In contrast to \acrshort{fDDA}, \acrshort{CDA}
additionally authenticates
the card's MAC \ACx{}.
The \SDADx{} is sent as GENERATE AC response, see~\secref{SubSec:Protocol:OTA}.

\subsubsection{Cardholder Verification}
The cardholder's presence is verified
using one of the \acrfullpl{CVM}: Paper Signature, Online \acrshort{PIN}, no \acrshort{CVM}, and \acrfull{CDCVM}.
For Paper Signature, the cardholder 
signs
the receipt, either on paper or on a touch screen. 
The cashier verifies the signature against the one on the card.
For Online PIN, the cardholder enters
a \acrshort{PIN} into the terminal,
which
encrypts the \acrshort{PIN} 
and sends it to the issuer for verification.
\acrshort{CDCVM} is 
performed by 
mobile devices such as smartphones 
by, e.g., verifying a PIN, a fingerprint, or a face scan.

Contactless transactions permit the use of no \acrshort{CVM} for \emph{low-value} transactions below a given limit.
\emph{High-value} transactions above this limit require some \acrshort{CVM}. 
There can also be a limit above which contactless transactions are forbidden and a contact transaction is required. 
These limits vary depending on the country and payment network, 
and change over time. 
Visa and Mastercard use different mechanisms to decide which \acrshort{CVM} is chosen, see \secreftwo{SubSec:Protocol:Mastercard}{SubSec:Protocol:Visa}.

\begin{figure*}[tb]  
    \centering
    \input{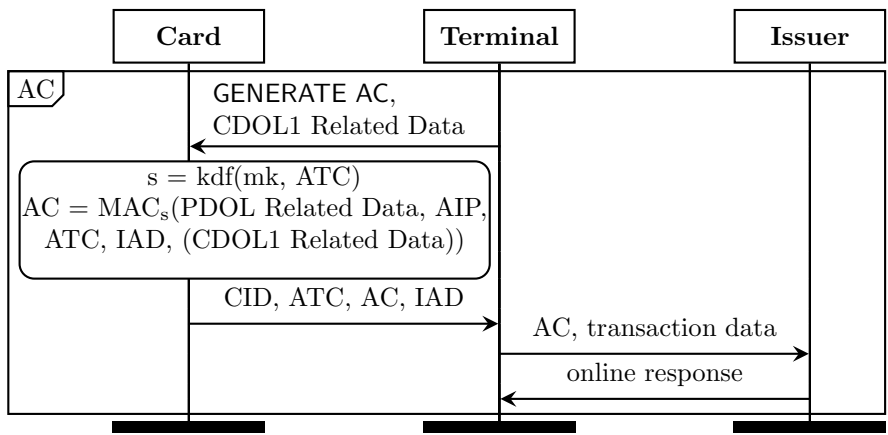}
    \caption{Overview of the Online Transaction Authorization phase. }
    \label{figure:Protocol:msc:TA}
\end{figure*}

\subsubsection{Online Transaction Authorization \textmd{(Figure~\ref{figure:Protocol:msc:TA})}}
\label{SubSec:Protocol:OTA}

The card authenticates 
to the issuer using a MAC, called \ACfull{}. 
The card sends it as 
a response to GENERATE AC
and the terminal forwards it 
to the issuer.

\looseness=-1
The card uses a symmetric ``master key'' \mkx{}, shared between the card and the issuer, and the \ATCfull{} to derive a session key. 
The \ATCx{} is incremented for each transaction 
and it is authenticated by the \ACx{} and depending on the application also by the signature \SDADx{}, see \secref{SubSec:Protocol:ODA}. 
Using this session key, the card 
computes the MAC \ACx{} over the transaction data. 
Note that the terminal does not have access to the key \mkx{} and thus cannot check the \ACx{}. 

\looseness=-1
A transaction can be declined or authorized either offline by the terminal or online by the issuer.
Offline authorization is not to be confused with \emph{\acrfull{ODA}}, described in \secref{SubSec:Protocol:ODA}.
The terminal requests online authorization
by sending the transaction data and the \ACx{}.
The issuer decides whether to authorize a transaction based on this data using proprietary fraud detection mechanisms~\cite{basinInducingAuthenticationFailures2023}.

\subsection{Visa and Mastercard}\label{SubSec:Protocol:MastercardAndVisa}
We now highlight features of Mastercard's kernel~2 and Visa's kernel~3 and 
explain how they differ compared to the above general protocol phases. 
We focus on these kernels as 
they are the targets of most reported attacks. 

\subsubsection{Mastercard \textmd{(Appendix~\ref{Appending:MSC}, Figure~\ref{figure:Protocol:Mastercard})}}\label{SubSec:Protocol:Mastercard}
The Mastercard kernel (kernel~2) mostly resembles the 
generic 
phases described in \secref{Section:DetailsOnEMV}.
It selects the \acrshort{CVM} according to the \AIPx{}, the \CVMListx{}, and the kernel configurations. 
The terminal informs the card of the chosen \acrshort{CVM} in the \CVMResultsx{}.
First, the terminal checks if \acrshort{CDCVM} can be performed, by checking the 
\AIPx{} and kernel configuration.
If they indicate that both the card and terminal support \acrshort{CDCVM}, the terminal checks if the transaction amount is above the \acrshort{CVM} required limit.
For low-value transactions, the \CVMResultsx{} indicate that no \acrshort{CVM} was performed and for high-value transactions the \CVMResultsx{} indicate that \acrshort{CDCVM} was performed.
If \acrshort{CDCVM} is not supported, the terminal checks if the \AIPx{} indicates that cardholder verification is supported.
If it is not supported, the \CVMResultsx{} indicate that no \acrshort{CVM} was performed.
Otherwise, the \acrshort{CVM} is chosen according to the \CVMListx{}, which indicates the card's supported \acrshortpl{CVM} and their priorities.
The terminal includes the \CVMResultsx{} in the \CDOLOneDatax{} in the GENERATE AC command.

Kernel~2's \acrfull{RRP} protects against relay attacks.
The card and terminal exchange timing data and nonces,
which are 
included in
the \SDADx{} signature.
\acrshort{RRP} seems to be not yet 
available in 
cards or terminals~\cite{raduPracticalEMVRelay2022a}.

\subsubsection{Visa \textmd{(Appendix~\ref{Appending:MSC}, Figure~\ref{figure:Protocol:Visa})}}\label{SubSec:Protocol:Visa}
The Visa kernel (kernel~3) is optimized to transmit fewer command/response pairs such that the card can be quickly removed from the terminal.
Thus, the Visa kernel omits the GENERATE AC command and sends the MAC \ACx{} in the GET PROCESSING OPTIONS response.
Visa also usually performs either \acrshort{fDDA} or online authorization, but rarely both.

\looseness=-1
Kernel 3 uses the \TTQfull{} and \CTQfull{} to 
choose a \acrshort{CVM}.
In particular, \TTQx{} indicates to the card the supported \acrshortpl{CVM} and whether a \acrshort{CVM} is required.
The \CTQx{} indicates to the terminal 
which \acrshortpl{CVM} the card requires
or if it performed \acrshort{CDCVM}.

\subsection{Mobile Devices}\label{SubSec:Protocol:Phone}
\looseness=-1
Mobile devices differ from physical, plastic cards as they can perform cardholder verification, called \acrshort{CDCVM}.
This is done, for example, 
using the cardholder's PIN, fingerprint, or face scan.
Apple Pay always requires 
\acrshort{CDCVM},
while Google Pay allows for low-value transactions without \acrshort{CDCVM}.

Note that Visa and Mastercard differ in how the terminal chooses \acrshort{CDCVM}: 
Mastercard uses the \AIPx{}, while Visa uses the the \CTQx{}.
\cite{raduPracticalEMVRelay2022a} observed that for Visa and Mastercard, the proprietary \IADfull{} indicates whether \acrshort{CDCVM} was performed and 
if a plastic card or a phone performed the transaction.
The \IADx{} is authenticated by the MAC \ACx{}, which 
Mastercard checks,
whereas Visa does not.

\paragraph{\textbf{Transit Mode}}
Apple Pay's ``Express Transit'' and Samsung Pay's ``Transport card'' are public transit features, called \emph{Transit Mode} by~\cite{raduPracticalEMVRelay2022a}.
Transit mode allows cardholders to present their phone at an origin and destination station to a dedicated transit terminal.
The transit providers can thereby record the route taken and compute and charge their fee.
The transit transaction is initially of value zero (and the actual cost is charged later)
and for convenience it does not require 
a \acrshort{CVM}.
Google Pay does not offer a dedicated transit mode as it already allows for low-value transactions without 
\acrshort{CVM}.

\section{Security Properties}\label{Section:SecurityProperties}
In this section, we describe the security properties that EMV should ideally achieve.
We derive these properties from the stakeholder's requirements, which we list first.
We then formulate these requirements more precisely as system properties.
We will list for each attack the violated properties in \secref{Section:AttacksOnEMV} and in Table~\ref{table:allAttacks}, in Appendix~\ref{Appendix:Tables}.
Note that most of the listed publications do not state which properties their attacks violate.

\subsection{Stakeholder Requirements}\label{SubSec:Properties:StakeholderRequirements}
Clearly, no stakeholder wants to lose money.
Some stakeholders have additional requirements.
In more detail, 
the cardholder requires the payment system to be available to pay with their card.
The cardholder does not want to lose money from unintended payments, so only intended transactions should be performed.
Finally, cardholders might desire privacy protection as card transactions can reveal personal data.

\looseness=-1
Merchants rely on 
the availability of the payment system.
Moreover, when a transaction is accepted by the terminal, the payment should succeed at the bank so that the merchant 
receives its payment and for the correct amount.

The acquirer, payment network, and issuer's business involves providing a secure, available payment system to the cardholders and merchants,
keeping their trust by fulfilling their requirements.
This in turn entails business decisions and risk analysis.
For example, it may be advantageous 
to incorrectly accept some payments when this reduces the chance of rejecting correct transactions.
Although wrongly accepted transactions can lead to financial losses, rejected transactions can drive away customers and thereby cause even greater losses due to reduced transaction volume.

\subsection{System Properties}\label{SubSec:SecProp:SystProp}

We now formulate the above abstract stakeholder requirements as system properties.
Note that some of these requirements go beyond the domain of EMV.
For example, the payment transfer from the cardholder's bank account to the merchant's account is out of EMV's scope.
Three of these system properties, namely
P1) \emph{data authentication}, P2) \emph{no delayed decline}, and P5) \emph{secrecy} were formalized by~\cite{basinEMVStandardBreak2021a,basinCardBrandMixup2021}.
We derive the other properties from the above stakeholder requirements.

\paragraph{\textbf{P1) Data Authentication}}
The EMV protocol should ensure that payments are performed as intended:
the correct funds should be transferred from the right cardholder account to the right merchant account.
This constitutes an agreement property between the card, terminal, and issuer:
They all should agree on the card's \PANx{} and the amount and currency of the transaction.
The agents might also want to agree on additional data to rule out unintended protocol flows,
like replay attacks which could lead to unintended (repeated) transactions.

\paragraph{\textbf{P2) No Delayed Decline}} \label{Subsection:SecurityProperties:NoDelayedDecline}
The issuer should not decline transactions that were accepted by the terminal.
Otherwise, a terminal could accept a transaction that might later be declined by the issuer, while the customer 
is long gone
with the goods
and the merchant would not get paid for these goods.

\paragraph{\textbf{P3) Cardholder Intent}}
Only transactions intended by the cardholder should be performed.
This is an abstract property as intent itself is difficult to formalize.  
Cardholder intent is approximated by two properties:

\textbf{P3.1) High-Value Transactions Require \acrshort{CVM}.}
\acrshortpl{CVM} increase confidence in the cardholder's intent. 
Thus, a \acrshort{CVM} should be performed for high-value transactions.

\textbf{P3.2) Card Close to Terminal.}
The card's proximity to the terminal can also imply cardholder intent.
Provided the card was not stolen, we can assume that the cardholder would only put the card close to the terminal if they intend to perform a payment. 
However, communication over NFC can be relayed and EMV contactless currently does not provide effective counter measures against it, see~\secreftwo{SubSec:Adversary:Practice}{SubSec:Flaws:Relay}.
Thus, this property is violated by standard \acrshort{MITM} attacks.
In fact, such \acrshort{MITM} attacks are an essential building block for many of the presented attacks.
Thus, we do not state which attacks violate this property in~\secref{Section:AttacksOnEMV}, but list them in Table~\ref{table:allAttacks} in Appendix~\ref{Appendix:Tables}.

\paragraph{\textbf{P4) Privacy}}
The cardholder might desire privacy 
regarding
personal data and metadata that might reveal 
their
habits.
Previous versions of the EMV protocol did not have any privacy protection mechanisms in place.
The new kernel~8 and the new Book~E~\cite{emvcoEMVContactlessSpecifications2023i} 
introduce 
such mechanisms.

\paragraph{\textbf{P5) Secrecy}}
Some of the EMV protocol's data should remain secret
to prevent fraudulent EMV contactless transactions.
This data includes the card's key material and PIN.
Other data can be misused in other ways such as 
forging magnetic stripe cards, see~\secref{SubSec:Flaws:MagStripeDataInEMV},
or in
card-not-present transactions.
This data includes the \PANx{}, expiry date, and \CSCfull{}.
Note that the \CSCx{} is printed on the card and it is called by VISA the \acrfull{CVV} and by Mastercard the \acrfull{CVC}.

\paragraph{\textbf{P6) Availability}}
All parties require the system's availability.

\section{Adversary Model}\label{Section:AdversaryCapabilities}

We next present a family of adversary models under which EMV's security can be studied.
We first present eight adversary capabilities and then highlight those seen in practice. 

This SoK focuses on the EMV contactless protocol.
Thus, we only consider adversary capabilities that target the protocol and its participants directly.
We do not consider attack surfaces outside of EMV contactless, such as the magnetic stripe or the banking system and we do not consider techniques not involving the protocol such as social engineering or ATM skimming.

\subsection{Adversary Capabilities}\label{SubSec:Adv:Cap}
We consider the four protocol participants:
the card, terminal, acquirer, and issuer, who communicate over three channels:
\begin{compactdesc}
    \item[\acrshort{NFC} channel:] between the card and the terminal;
    \item[Acquirer network:] between the terminal and the acquirer;
    \item[Payment network:] between the acquirer and the issuer;
\end{compactdesc}

\setlength{\multicolsep}{0pt plus 1.0pt minus 0.75pt}
We consider three adversary capabilities. 
(1)~\emph{Network capabilities}: The adversary can control some of the three channels.
(2)~\emph{Compromising capabilities}: The adversary can compromise 
some of the agents, thereby learning their secret keys.
This lets the adversary impersonate this agent.
(3)~\emph{Visual channel capabilities}: The adversary 
can 
visually inspect a card.
This results in the following eight combinations: 
\renewcommand{\labelenumi}{A\arabic{enumi})}
\begin{multicols}{2}
\begin{enumerate}[topsep=0pt,itemsep=-1ex]
    \item control \acrshort{NFC} channel,
    \item control acquirer netw.,
    \item control payment netw.,
    \item compromise card,
    \item compromise terminal,
    \item compromise acquirer,
    \item comp. issuer, and
    \item access visual channel.
\end{enumerate}
\end{multicols}
\noindent Any subset of these combinations yields an adversary model.

\subsection{Adversary in Practice} \label{SubSec:Adversary:Practice}
We assess the adversary capabilities as seen in practice.

\paragraph{\textbf{A1) \acrshort{NFC} Channel}}\label{SubSec:Adversary:NFCChannel}
\looseness=-1
The 
NFC interface of EMV contactless is more easily accessed by the adversary than the card-terminal channel of EMV contact.
The NFC channel can be accessed from a short distance without direct contact to the card or terminal.
The adversary can eavesdrop on the communication 
on the NFC channel.
The adversary can also communicate with the card or terminal by emulating a card or terminal.

NFC's limited range can be extended by relay attacks, which have been demonstrated on EMV contactless.
The relay infrastructure usually consists of two devices communicating over a relay channel. 
One device, the terminal emulator, communicates to the card and the other, the card emulator, communicates to the terminal.
This infrastructure can be implemented using two NFC enabled Android phones~\cite{francisPracticalGenericRelay2013, emmsDangersVerifyPIN2012, chothiaRelayCostBounding2015} 
or specialized hardware~\cite{shanManNFC2017}.
A \acrshort{MITM} attack can be performed by modifying the relayed messages.
Note that a relay attack can be seen as an attack 
violating the card's proximity to the terminal~(P3.2)
rather than an adversary capability. 
We categorize it as an adversary capability, as it is a central building block of most of the attacks described in this SoK.

\paragraph{\textbf{A2, A3) Acquirer and Payment Networks}}

The EMV specification~\cite{emvcoEMVIntegratedCircuit2022d} specifies the acquirer interface.
However, communication between the acquirer and the issuer is not part of this specification.
We are unaware of practical attacks targeting the acquirer or payment networks.

\paragraph{\textbf{A4) Compromise Card}}
\looseness=-1
The extraction of key material through side channels, such as timing attacks~\cite{kocherTimingAttacksImplementations1996, dhemPracticalImplementationTiming2000} or power analysis attacks~\cite{PowerAnalysisAttacks2007}, or using invasive methods such as microprobing~\cite{koemmerlingDesignPrinciplesTamperResistant1999} and its countermeasures are widely studied and still an active research area.
Smart card manufacturers continue to deploy new countermeasures, while new attacks are being developed.
Extracting the card's key material allows an adversary to impersonate the card by forging the MACs and signatures.
This allows the adversary to forge transactions.
As the adversary requires access to the card to extract the key materials, the benefit is unclear of extracting the key material over just using a stolen card.

\paragraph{\textbf{A5) Compromise Terminal}}
\looseness=-1
EMV does not specify any of the terminal's secret key material.
It might however possess keys to communicate with the acquirer.
In addition to revealing key material, a terminal could be modified.
The viability of such modifications was demonstrated on a live system by~\cite{gallowayFirstContactNew2019}, 
who 
modified the terminal's
nonce choice, see~\secref{SubSec:Attacks:ReplayUNReuse}.

\looseness=-1
The attack by~\cite{emmsHarvestingHighValue2014}, see~\secref{SubSec:Attacks:ForeignCurrency}, relies on a rogue terminal that forwards a recorded transaction.
\cite{emmsHarvestingHighValue2014}~implemented a proof-of-concept rogue terminal.
They did not test it on a live system
and thus it is unclear how practical this attack is.

Other terminal modifications include
displaying
a wrong payment amount to overcharge the cardholder, or recoding the entered PIN.
The feasibility of arbitrary modifications was demonstrated by~\cite{ChipPINTerminal2006}.
Such modifications would most likely need to be performed by the merchant, as the terminal providers employ physical tamper resilience. 
Note that if 
such modifications are detected, 
the merchant owning this terminal is not anonymous as the merchant has an account with the acquiring bank.
This makes such attacks unattractive.

Note that a \emph{fake} terminal relaying a transaction to a \emph{real} terminal does not constitute a compromised terminal but is part of a \acrshort{MITM} infrastructure on the NFC channel~(A1).

\paragraph{\textbf{A6, A7) Compromise Acquirer or Issuer}}
The acquirer and issuer have control over the merchant's and cardholder's accounts respectively. 
Thus,
an attacker controlling an acquirer or issuer can perform more powerful attacks by accessing these accounts instead of attacking the EMV protocol.

\paragraph{\textbf{A8) Visual Adversary}}
An adversary may visually inspect a card.
Whenever a plastic card 
is used for a payment, it 
is exposed and an adversary can potentially see its details.
The installation of inconspicuous cameras near terminals, as proposed by~\cite{emmsPracticalAttackContactless2011}, is also possible.
The adversary could also observe the cardholder entering the PIN into a terminal.

\section{Flaws}\label{Section:Flaws}

In this section, we list the flaws identified in the EMV contactless protocol.
As we will see in~\secref{Section:AttacksOnEMV}, some flaws are combined into practical attacks.
In~\secref{Section:Discussion:Flaws} we will discuss the flaws listed in here and highlight some common traits.

We categorize each flaw with respect to 
the protocol variant where it appears, namely the Mastercard or Visa kernel, or Mobile Devices, and when  the flaw is not specific to a protocol variant, we categorize it under ``General EMV''.
We list flaws on mag-stripe mode in Appendix \secref{SubSection:Flaws:MagStripeMode}.

\subsection{General EMV}
We now list the flaws not specific to  any protocol variant.

\subsubsection{No Relay Protection}\label{SubSec:Flaws:Relay}
\looseness=-1
NFC communication 
can be relayed, which is exploited by \acrshort{MITM} attacks.
Multiple protocols to provide relay protection have been proposed~\cite{drimerKeepYourEnemies,raduPracticalEMVRelay2022a, chothiaRelayCostBounding2015}, also by
Visa~\cite{chenBindingCryptogramProtocol2022a}
and Mastercard, see~\secref{SubSec:Protocol:Mastercard}.
However, to our knowledge, these protocols are not yet
available
in cards or terminals.
Moreover,~\cite{raduPracticalEMVRelay2022a} 
demonstrated the ineffectiveness of these protocols.

\subsubsection{Visa Responses Built from Mastercard Responses}\label{SubSec:Flaws:VisaFromMastercard}
A terminal communicating with a Visa card should run the Visa kernel and a terminal communicating with a Mastercard card should run the Mastercard kernel.
However,~\cite{basinCardBrandMixup2021} showed that responses for the Visa kernel can be built from the responses of a Mastercard card.
This enabled the \emph{Card Brand Mixup} attack by~\cite{basinCardBrandMixup2021}, see \secref{SubSec:Attacks:PINBypassMastercard:CardBrandMixup}.
This attack is possible for the insecure Visa configuration without \acrshort{fDDA}. 
Whereas transactions with \acrshort{fDDA} prevent this attack, 
as Visa's \SDADx{} has a header different than Mastercard's. 
A Mastercard \SDADx{} will therefore not be accepted by the Visa kernel.

\subsubsection{Not Checked if \AIDx{} and \PANx{} Match}\label{SubSec:Flaws:CheckAIDPAN}
Terminals communicating with a card should be running a kernel corresponding to the card's brand.
The card's brand can be determined from the \PANx{}, and the \AIDx{} indicates which kernel the terminal is running.
Thus, issuers could verify that the \PANx{} and \AIDx{} match.
However, 
experiments by~\cite{basinCardBrandMixup2021} and their disclosure process with Mastercard 
revealed that not all issuers 
detect mismatches.
As a result of this process,
Mastercard implemented those checks on their network.
Note that the behavior of the issuers is not publicly specified.

\subsubsection{TAC-Denial Set to Zero}\label{SubSec:Flaws:TACDenail}
The \TACDenialfull{} specifies the acquirer's conditions that cause transactions to be declined offline.
The terminal checks the \TACDenialx{} and \IACdenialx{} against the \TVRfull{} 
to determine if a transaction should be declined.
The specification recommends not to set some Terminal Action Codes to zero.\footnote{According to~\cite{emvcoEMVIntegratedCircuit2022d} Book 3, page 123, specific bits, including the `CDA Failed' bit, of the \TACOnlinex{} and \TACDefaultx{} should be set to one.} However, there are no recommendations for the \TACDenialx{}.
The experiments conducted by~\cite{basinInducingAuthenticationFailures2023} showed that many terminals set the `CDA Failed' bit in the \TACDenialx{} to zero. 
This prevents transactions with this bit set from being declined, 
which allows fraudulent transactions to be performed, such as the
PIN-Bypass attack by~\cite{basinInducingAuthenticationFailures2023}, see \secref{SubSection:Attacks:InducingAuthenticationFailure}.

\subsubsection{Unencrypted Data}\label{SubSec:Flaws:UnencryptedData}

Some data is transmitted unencrypted in EMV transactions, and can be misused by the adversary.
For example, the \emph{Magnetic Stripe Cloning} attack~\cite{gallowayItOnlyTakes2020} in \secref{SubSec:Attacks:MagStripeClone} exploits the unencrypted \TrackTwoeqdatax{} and the \emph{Eavesdropping on Card Data} attack~\cite{emmsPracticalAttackContactless2011} in \secref{SubSec:Attacks:EavesdroppCardData} harvests the unencrypted \PANx{}, cardholder name, and expiry date.

\subsubsection{\acrshort{SDA} and \acrshort{DDA} do Not Authenticate the \ACx{}}\label{SubSec:Flaw:ACNotAuth}

\acrshort{ODA} authenticates the card to the terminal. 
The \acrshort{ODA} methods \acrshort{SDA} and \acrshort{DDA} only authenticate card data but not transaction data such as the MAC \ACx{}.
\cite{basinEMVStandardBreak2021a} showed that the adversary can modify the \ACx{}, which results in the \emph{Merchant-Holding-the-Bag} attack described in \secref{SubSec:Attacks:MerchantHoldingBag}.

\subsubsection{Weakness of Paper Signatures}\label{SubSec:Flaws:SignatureInterface}
\looseness=-1
Paper signatures can be forged and cashiers might not properly check signatures.
Using paper signatures
is even worse when the transaction is performed with a smart device like a smartphone.
The cashier will use a signature presented on the phone's screen 
for verification.
When the phone is 
under the adversary's control,
it can show 
any signature.
This is exploited in the \emph{Inducing Authentication Failure} attack by~\cite{basinInducingAuthenticationFailures2023}, see \secref{SubSection:Attacks:InducingAuthenticationFailure}.

\subsubsection{Out-Of-Order \ATCxpl{} Accepted}\label{SubSec:Flaws:ATCoutOfOrder}
\looseness=-1
The \ATCx{} should prevent the authorization of out-of-order transactions.
However,~\cite{gallowayFirstContactNew2019} showed that many issuers do not properly check the \ATCx{}.
Their experiments 
showed that transactions 
were accepted with an out-of-order \ATCx{}.
These checks are not part of the EMV specification.

\looseness=-1
\subsubsection{UN Reuse Not Prevented}\label{SubSec:Flaws:NonceReuse}\label{SubSec:Flaws:UNReuse}
The terminal-sourced nonce \UNTx{} should prevent replay attacks,
but \cite{gallowayFirstContactNew2019} showed that 
terminals can
resend the \UNTx{},
enabling the \emph{Replay with Nonce Reuse} attack, 
see 
\secref{SubSec:Attacks:ReplayUNReuse}.

\subsubsection{\AIDx{} is Not Cryptographically Protected}\label{SubSec:Flaws:AIDNotProtected}
The card informs the terminal of its supported kernels by sending \AIDxpl{} and the terminal chooses a kernel by sending the corresponding \AIDx{}.
Neither the \AIDxpl{} sent by the card nor the terminal are cryptographically protected and hence they can be modified by an attacker.
This enables the \emph{Card Brand Mixup} attack by~\cite{basinCardBrandMixup2021} and the 
downgrade attack by~\cite{basinInducingAuthenticationFailures2023}, see \secreftwo{SubSec:Attacks:PINBypassMastercard:CardBrandMixup}{SubSec:Attacks:Downgrade:MaestroToMastercard}.

\subsubsection{Not Checked if Plastic Cards Perform \acrshort{CDCVM}}\label{subsection:Flaws:9.29:cardsPerformCDCVM}\label{SubSec:Falaws:PlasticCardsCDCVM}
\acrshort{CDCVM} can only be performed by mobile devices and therefore not by
plastic cards.
However, according to~\cite{gallowayFirstContactNew2019}, the card's type and the performed \acrshort{CVM} are not checked.
Namely, 
declining
transactions with \acrshort{CDCVM} and an \IADx{} indicating a plastic card,
could prevent the PIN-bypass attacks by~\cite{gallowayFirstContactNew2019, basinEMVStandardBreak2021a,basinCardBrandMixup2021}, see \secrefthree{SUBSEC:Attacks:FirstVisaPINBypass}{SubSec:Attacks:2ndVisaPINBypass}{SubSec:Attacks:PINBypassMastercard:CardBrandMixup}.

\subsubsection{Magnetic Stripe Data Present in EMV}\label{SubSec:Flaws:MagStripeDataInEMV}

Magnetic stripe payment cards mostly consist of two data tracks,
containing
persistent data such as the \PANx{} and the expiration date.
\cite{gallowayItOnlyTakes2020} demonstrate that 
these tracks can be constructed from EMV contactless transactions.
The EMV \TrackOneTwoeqdatax{} are almost identical to the magnetic stripe tracks. 
It can be accessed by the READ RECORD command in EMV mode and is still present in the latest kernel specifications and the new kernel~8.

The magnetic stripe tracks contain a static \CSCfull{}, which authenticates the card. 
Compared to 
this 
static \CSCx{},
the \TrackTwoeqdatax{} in EMV contains a dynamic \CSCx{}. 
This, and the fact that the discretionary data should be unique for each transaction mode, should prevent such a cloning attack. 
However, experiments by~\cite{gallowayItOnlyTakes2020} show that these measures are not properly implemented.

\subsubsection{Magnetic Stripes Are Still Supported}\label{SubSec:Flaws:MagStripeStillSupported}

Most EMV cards still contain a magnetic stripe 
that is
vulnerable to cloning.
The data can be read with a publicly available reader and written to a new magnetic stripe card. 
This makes the magnetic stripe the weakest link on the card. 
Mastercard announced that they would remove magnetic stripes from future cards~\cite{SwipingLeftMagnetic}, but they are still there in 2025.
Although this is not directly a flaw of the EMV contactless protocol, it makes attacks where data is extracted from an EMV transaction impactful, see~\secreftwo{SubSec:Flaws:MagStripeDataInEMV}{SubSec:Attacks:MagStripeClone}.

\subsection{Mastercard Kernel}

We next list the flaw discovered in the Mastercard kernel.

\subsubsection{CA Look up Failure Not Declined}\label{SubSub:Flaws:ODAFailure}\label{SuSec:Flaws:CALookupFailure}
One would expect a terminal to decline transactions when \CApkx{} lookup fails.
However,~\cite{basinInducingAuthenticationFailures2023} showed that the Mastercard kernel does not always decline such transactions.

According to the 
Mastercard kernel 
specification~\cite{emvcoEMVContactlessSpecifications2023i},
the terminal sets the `CDA Failed' bit of the \TVRx{} when the \CApkIndexx{} is not present in the terminal's database. 
This bit is set before the GENERATE AC command and it directly affects whether \acrshort{CDA} is performed.
Namely, if the \TVRx{} has the `CDA Failed' bit set and the \AIPx{} indicates that the card does not support \acrshort{CDCVM} then the terminal does not request the \SDADx{} in the GENERATE AC command.
Thus, no \acrshort{ODA} will be performed and all data that is only protected by \acrshort{ODA} will be vulnerable to modifications.

\subsection{Visa Kernel}
We next list the flaws discovered in the Visa kernel.

\subsubsection{CTQ is Not Protected}\label{SubSec:Flaws:CTQnotProtected}
In Visa transactions, the \CTQx{} tells the terminal the card's \acrshort{CVM} requirements and capabilities.
The \CTQx{} is authenticated through the signature \SDADx{}. 
However, the \CTQx{} is not protected for transactions without \acrshort{fDDA}.
This flaw was discovered by~\cite{gallowayFirstContactNew2019} and exploited in the PIN-bypass attacks by~\cite{gallowayFirstContactNew2019, basinEMVStandardBreak2021a,raduPracticalEMVRelay2022a,basinCardBrandMixup2021}, see \secreffour{SUBSEC:Attacks:FirstVisaPINBypass}{SubSec:Attacks:2ndVisaPINBypass}{SubSec:Attacks:PINBypassMastercard:CardBrandMixup}{SubSec:Attacks:MagicByte}.

\subsubsection{No Limit for Foreign Currencies}\label{SubSec:Flaws:ForeignCurrency}
\cite{emmsHarvestingHighValue2014} discovered that many Visa credit cards 
do not require a \acrshort{CVM} for
high-value transactions in a foreign currency.
This allows the \emph{No Cardholder Veriﬁcation in Foreign Currencies} attack, see \secref{SubSec:Attacks:ForeignCurrency}.

\subsubsection{PIN Verification Over \acrshort{NFC}}\label{SubSec:Flaws:OfflinePINoverNFC}
Encrypted Offline \acrshort{PIN} is intended for EMV contact. 
The terminal encrypts the \acrshort{PIN} and sends it to the card, which verifies the PIN.
In contrast, in contactless transactions, the \acrshort{PIN} is verified by the issuer via Online \acrshort{PIN}.
\cite{emmsDangersVerifyPIN2012, emmsRisksOfflineVerify2013} showed that the Offline \acrshort{PIN} functionality is available over \acrshort{NFC},
also offering queries
for the remaining \acrshort{PIN} attempts.
\cite{emmsDangersVerifyPIN2012, emmsRisksOfflineVerify2013} exploit 
this
in their \emph{PIN-Guessing} attacks, see \secref{SubSec:Attacks:PINGuess}.
Offline PIN was available
on Visa cards 
but not on
the tested Mastercard cards.

\subsubsection{Merchant Details are Not Authenticated}\label{SubSec:Flaws:MerchantNotAuth}
The card does not authenticate any merchant or terminal data with the MAC \ACx{}.
Thus, the issuer determines the merchant account based on the data from the acquirer.

\subsubsection{\acrshort{TTQ} is Not Protected}\label{SubSec:Flaws:TTQnotProtected}
In VISA transactions, the \TTQx{} informs the card of the terminal's \acrshort{CVM} capabilities and requirements.
Experiments by~\cite{gallowayFirstContactNew2019} showed that the \TTQx{} could be modified.
According to \cite{raduPracticalEMVRelay2022a}, it is unclear whether the \TTQx{} is omitted from the \ACx{} or if the issuer does not check the \TTQx{}.
This flaw is exploited by the PIN-bypass attacks 
by~\cite{gallowayFirstContactNew2019, raduPracticalEMVRelay2022a}, see \secreftwo{SUBSEC:Attacks:FirstVisaPINBypass}{SubSec:Attacks:MagicByte}.

\subsection{Mobile Devices}
We next list the flaws in mobile devices
such as phones.

\subsubsection{Mobile Devices Always Send \acrshort{CDCVM}}\label{SubSub:Flaws:SendingCDCVM}
Google Pay allows for low-value transactions 
with a locked phone.
In contrast, high-value transactions require 
the phone to be unlocked.
In the Visa kernel, the card indicates with the \CTQx{} if \acrshort{CDCVM} was performed.
\cite{gallowayFirstContactNew2019} observed that Google Pay Visa cards always send a \CTQx{} indicating that \acrshort{CDCVM} was performed if the card accepts the transaction,
independently of whether the screen was locked.
This is exploited in the PIN-bypass attack on Google Pay, 
see \secref{SubSec:Attacks:FirstVisaPINBypassGooglePay}

\subsubsection{Transit Mode Initiated with ``Magic Byte''}\label{SubSec:Flaws:MagicByte}
\cite{raduPracticalEMVRelay2022a} discovered that transit terminals send some ``magic bytes'' to unlock Apple Pay and initiate a transit mode transaction. 
These ``magic bytes'' can be recorded and replayed.
\cite{raduPracticalEMVRelay2022a} exploits this in their PIN-Bypass attack,
see \secref{SubSec:Attacks:MagicByte}.

\section{Attacks on EMV}\label{Section:AttacksOnEMV}
In this section, we show how the flaws just presented are exploited by surveying the main attacks on EMV contactless reported in the literature.
For each attack, we state the violated security properties~(P1-P6) and the required adversary capabilities~(A1-A8), as presented in~\secref{Section:SecurityProperties} and \secref{Section:AdversaryCapabilities},
which
we also list in
Table~\ref{table:allAttacks} in Appendix~\ref{Appendix:Tables}.
As each attack requires 
controlling the NFC channel~(A1), 
we will not list this capability here.

We structure the attacks according to seven classes: card cloning, replay, PIN-guessing, denial of service, PIN-bypass, merchant-holding-the-bag, and downgrade attacks.
We further structure the PIN-bypass attacks into attacks on Visa, Mastercard, and mobile devices.
In Appendix \secref{SubSec:Attacks:MagStripeMode}, we describe the attacks on mag-stripe mode.

\subsection{Card Cloning}
In a card cloning attack, a physical card is created
that mimics the target card and is accepted in transactions.

\subsubsection{Magnetic Stripe Cloning}\label{SubSec:Attacks:MagStripeClone}
\looseness=-1
The attack by~\cite{gallowayItOnlyTakes2020} extracts data from EMV mode transactions to create a magnetic stripe card. 
An adversary 
records the unencrypted \TrackOneTwoeqdatax{}, see \secref{SubSec:Flaws:UnencryptedData}, and extracts the magnetic stripe data from it, see \secref{SubSec:Flaws:MagStripeDataInEMV}.
This data is written 
on a magnetic stripe card, 
forging a 
magnetic stripe card for the 
victim's account.
This attack is
viable as the magnetic stripe is still supported, see \secref{SubSec:Flaws:MagStripeStillSupported}, and as 
magnetic stripes
mostly use the insecure paper signature, see \secref{SubSec:Flaws:SignatureInterface}. 
This attack extracts and misuses data from an EMV contactless transaction, 
violating the secrecy property~(P5).
This attack does not violate further properties,
as it does not forge an EMV contactless transaction but a magnetic stripe transaction.

\subsection{Replay Attacks}
In a replay attack, also known as skimming, eavesdropping, or pre-play, transaction data is recorded and 
reused later.

\subsubsection{No Cardholder Verification in Foreign Currencies}\label{SubSec:Attacks:ForeignCurrency}
\looseness=-1
The attack by~\cite{emmsHarvestingHighValue2014} exploits the lack of cardholder verification for foreign currencies, see \secref{SubSec:Flaws:ForeignCurrency}.
The authors exploit this flaw in a replay attack.
The adversary records a transaction between a terminal emulator 
and a target card.
Afterwards, the transaction is sent to a rogue merchant~(A5) who sends the transaction to the acquirer who forwards it to the issuer.
If the issuer accepts the transaction, the funds are transferred from the victim cardholder's bank account to the rogue merchant's account.
As the card does not authenticate the merchant details to the issuer, see  \secref{SubSec:Flaws:MerchantNotAuth},
the recorded transaction data could be used by any rogue merchant.
This attack violates the cardholder's intent~(P3) and bypasses the \CVM{}~(P3.1).

\cite{emmsHarvestingHighValue2014}
implemented a proof-of-concept rogue merchant,
that was not tested
against a live 
system. 
Hence, the attack's practicality is unclear.
It is unclear too 
if this flaw could be exploited in a \acrshort{MITM} attack
or if it would be more practical to 
directly use a stolen card abroad.
\cite{emmsHarvestingHighValue2014} argues for 
a replay attack
as
no synchronization 
is required, in contrast to a \acrshort{MITM} attack.

\subsubsection{Replay with Nonce Reuse}\label{SubSec:Attacks:ReplayUNReuse}
\looseness=-1
\cite{gallowayFirstContactNew2019}
record a transaction between a card and a terminal emulator
for 
a given \UNTx{}. 
This transaction can then be replayed to a modified terminal~(A5) that always uses this \UNTx{}. 

\looseness=-1
The fresh nonce \UNTx{} from the terminal and the counter \ATCx{} from the card should prevent replay attacks. 
However,~\cite{gallowayFirstContactNew2019} demonstrated that the issuer does not check if a terminal uses the same \UNTx{} multiple times and that some issuers accept out-of-order \ATCxpl{}, see \secreftwo{SubSec:Flaws:ATCoutOfOrder}{SubSec:Flaws:NonceReuse}. 
This attack violates 
the cardholder's intent~(P3)
and
the stronger notions of data authentication~(P1) required to rule out replay attacks.

\subsection{PIN Guessing}\label{SubSec:Attacks:PINGuess}

In the \emph{PIN-Guessing} attacks by~\cite{emmsDangersVerifyPIN2012, emmsRisksOfflineVerify2013},
the adversary
automatically guesses and tests different \acrshortpl{PIN} on cards. 
A ``guessing device'' is placed at an entrance that requires NFC access cards.
The attack relies on 
victims
placing their entire wallet at the NFC reader 
and thus in NFC range of the guessing device.
Guessing the \acrshort{PIN} correctly violates the secrecy of the \acrshort{PIN} (P5).
An adversary could then steal the 
card and 
use it at 
a \acrshort{PoS} terminal,
violating 
the cardholder's intent (P3).

This attack exploits the Offline \acrshort{PIN} functionality 
over \acrshort{NFC}, see \secref{SubSec:Flaws:OfflinePINoverNFC}. 
The guessing device can 
check 
the remaining PIN attempts
to avoid blocking the card.
If no attempts are left,
the attack can be delayed until the cardholder resets
the remaining attempts
at a \acrshort{PoS} terminal or ATM.
\cite{emmsRisksOfflineVerify2013} 
extend this attack 
by
guessing the PINs of different cards in a wallet individually.

\subsection{Denial of Service Attacks}
A denial of service (DoS) attack prevents the correct usage of the payment system and violates availability~(P6).

\paragraph{\textbf{PIN-Guess-Spamming}}\label{Subsec:Attacks:PINSpamming}
\cite{emmsDangersVerifyPIN2012} propose using their \emph{\acrshort{PIN}-Guessing} attack as a DoS attack by querying \acrshortpl{PIN} until the attempt limit is reached, 
blocking
the card. 
The adversary's motivation could be to 
be a nuisance
or blackmail issuing banks by threatening to upset customers~\cite{emmsRisksOfflineVerify2013}.

\subsection{PIN-Bypass Attacks on Visa}\label{Subsec:Attacks:PINBypassVisa}
High-value transactions
require 
cardholder verification,
which is for physical cards 
mostly Online PIN. 
With a PIN-bypass attack however, the adversary can perform a high-value transaction without providing the PIN~(P3.1), which violates the cardholder's intent~(P3).
\cite{emmsDangersVerifyPIN2012} explain how PIN-bypass attacks could be performed without providing a concrete attack:
A \acrshort{MITM} attacker can modify control data such as the \acrshort{CVM}.

\subsubsection{Combined TTQ, CTQ Modification Attack}\label{SUBSEC:Attacks:FirstVisaPINBypass}
The first PIN-bypass attack on contactless transactions was described by~\cite{gallowayFirstContactNew2019}.
The attack 
modifies the \TTQx{} and \CTQx{},
which are used to choose the \acrshort{CVM}.
The adversary first modifies the \TTQx{} from the terminal to indicate that no cardholder verification is required. 
The adversary then modifies the \CTQx{} from the card to indicate that \acrshort{CDCVM} was performed.
As a result, the terminal will assume that it is talking to a phone that performed \acrshort{CDCVM} and it will not require \acrshort{PIN} verification.
Thus, the card and terminal disagree~(P1) on the \TTQx{} and \CTQx{}.

These modifications are possible as both the \TTQx{} and \CTQx{} are not cryptographically protected in Visa transactions without \acrshort{fDDA}, see \secreftwo{SubSec:Flaws:TTQnotProtected}{SubSec:Flaws:CTQnotProtected}.
\cite{gallowayFirstContactNew2019} and~\cite{raduPracticalEMVRelay2022a} point out that this attack could be detected 
and prevented
by checking that \acrshort{CDCVM} is not performed by plastic cards, see \secref{subsection:Flaws:9.29:cardsPerformCDCVM}.

\subsubsection{CTQ Modification Attack}\label{SubSec:Attacks:2ndVisaPINBypass}
The \acrshort{PIN}-bypass attack by \cite{basinEMVStandardBreak2021a} is 
similar to the attack by~\cite{gallowayFirstContactNew2019}. 
The attack also targets Visa transactions without \acrshort{fDDA}.
This attack only requires modifying the \CTQx{} and not the \TTQx{}. 
However, an additional bit of the \CTQx{} is modified, 
indicating 
to the terminal that Online \acrshort{PIN} verification is not required.

\subsection{PIN-Bypass on Mastercard}
\looseness=-1
The PIN-bypass attacks on Visa just described are not directly possible on Mastercard's kernel~2 as the \AIPx{}, which indicates the supported and required \acrshortpl{CVM}, is cryptographically protected through the \ACx{} MAC and, if present, the \SDADx{} signature.
In this section, we describe two PIN-bypass attacks that also violate cardholder intent~(P3) and bypass the PIN~(P3.1).

\subsubsection{Card Brand Mixup}\label{SubSec:Attacks:PINBypassMastercard:CardBrandMixup}
\looseness=-1
\cite{basinCardBrandMixup2021} 
trick the terminal 
into running the Visa kernel with a Mastercard card.
This lets the adversary 
to perform the \emph{CTQ Modification} PIN-bypass attack, see~\secref{SubSec:Attacks:2ndVisaPINBypass}, with a Mastercard card.

The card and terminal choose a kernel according to the \AIDxpl{} sent.
As the \AIDxpl{} are not cryptographically protected, see \secref{SubSec:Flaws:AIDNotProtected},
the adversary can modify the card-sourced \AIDxpl{} to 
indicate 
that the card only supports the Visa kernel.
The terminal will thus choose the Visa kernel and respond with
the Visa \AIDx{}.
The adversary replaces it with the Mastercard \AIDx{}, resulting in a disagreement~(P1) on the \AIDx{}.
The terminal therefore runs the Visa kernel while the card runs the Mastercard kernel. 
The adversary harvests the terminal's Visa commands to build Mastercard commands for the card and 
vice versa for the card's responses,
see \secref{SubSec:Flaws:VisaFromMastercard}.

Experiments by~\cite{basinCardBrandMixup2021} showed that 
this attack
is possible with some terminals,
whereas, other terminals decline modified transactions. 
The declined transactions were probably detected by the issuer using  
additional checks on the transaction data.
As part of the disclosure process of~\cite{basinCardBrandMixup2021}, Mastercard implemented measures on their network to mitigate this 
attack.
The \AIDx{} is checked against the \PANx{},
see \secref{SubSec:Flaws:CheckAIDPAN}. 
Experiments by~\cite{basinCardBrandMixup2021} showed 
the effectiveness of those measures.

\subsubsection{Inducing Authentication Failure}\label{SubSection:Attacks:InducingAuthenticationFailure}

\cite{basinInducingAuthenticationFailures2023} discovered another PIN-bypass attack on the Mastercard kernel that exploits authentication failures. 
The attack enables the modification of all data that is not authorized online through the \ACx{} by the issuer. 
This includes the \CVMListx{} that informs the terminal of the card's supported \acrshortpl{CVM}. 
The \acrshort{PIN} can be bypassed by either replacing the \CVMListx{} with an empty list or replacing Online \acrshort{PIN} with Paper Signature.
Using a phone as the card emulator enables the attacker to display their own signature on the screen, so the cashier can verify the attacker's signature, see \secref{SubSec:Flaws:SignatureInterface}. 

This authentication failure is induced by replacing the \CApkIndexx{} with an invalid value.
Thus, the terminal fails
to access the \CApkx{},
it 
does not request the \SDADx{}, see~\secref{SubSub:Flaws:ODAFailure}, 
and the \CVMListx{} is not authenticated.
The `CDA Failed' bit in the \TVRx{} is thus set. 
If this bit is also set in the \IACdenialx{} or \TACDenialx{}, the terminal declines.
However, the adversary can replace the \IACdenialx{} from the card 
and many terminals have this bit in the \TACDenialx{} set to zero, see \secref{SubSec:Flaws:TACDenail}.
This results in the terminal not declining and not authenticating the \CVMListx{} and thus accepting the modified \acrshortpl{CVM}.
Thus, the terminal and card disagree~(P1) on the \CApkx{}, \CVMListx{}, and \IACdenialx{}.

\subsection{CVM-Bypass on Mobile Devices}
Mobile devices
offer 
\acrshort{CDCVM}.
In a
\acrshort{CVM}-bypass attack,
an adversary pays
with a locked phone and bypasses 
\acrshort{CDCVM}~(P3.1),
violating 
cardholder intent~(P3).

\subsubsection{TTQ, CTQ Modification Attack on Google Pay}\label{SubSec:Attacks:FirstVisaPINBypassGooglePay}
The 
PIN-Bypass attack by~\cite{gallowayFirstContactNew2019}, see \secref{SUBSEC:Attacks:FirstVisaPINBypass}, could also be applied to Google Pay.
Low-value Google Pay transactions do not require \acrshort{CDCVM}
and they can be
performed with a locked phone.
In contrast to the attack on plastic cards, 
in the Google Pay attack  
only the \TTQx{} must be modified.
In the plastic card attack, the \CTQx{} is modified to indicate that \acrshort{CDCVM} was performed.
This modification is not required when attacking Google Pay as mobile devices always send \emph{\acrshort{CDCVM} performed} when they accept a transaction, see \secref{SubSub:Flaws:SendingCDCVM}.

\subsubsection{``Magic byte'' PIN-Bypass}\label{SubSec:Attacks:MagicByte}

\cite{raduPracticalEMVRelay2022a} 
exploit Apple Pay's transit mode and the Visa kernel. 
Replaying some ``magic bytes'' unlocks Apple Pay on a locked iPhone, see \secref{SubSec:Flaws:MagicByte}. 
This allows 
relaying a transit mode transaction between a initially locked iPhone and a transit terminal.

To relay a non-transit transaction to a non-transit terminal,~\cite{raduPracticalEMVRelay2022a} modified the unprotected \TTQx{}, see \secref{SubSec:Flaws:TTQnotProtected}, 
such that the \emph{\acrshort{ODA} for Online Authorizations supported} bit and the \emph{EMV mode supported} bit are set.
\cite{raduPracticalEMVRelay2022a} observed a low success rate due to the timings of the relayed messages.
The success rate could be improved by caching the READ RECORD response.

The above modifications enable the attacker to relay 
a low-value transaction between a non-transit terminal and a locked iPhone. 
The attack can be extended for high-value transactions similarly to the PIN-bypass attack on Visa cards by~\cite{basinEMVStandardBreak2021a}, see \secref{SubSec:Attacks:2ndVisaPINBypass}.
The adversary modifies the unprotected \CTQx{}, see \secref{SubSec:Flaws:CTQnotProtected}, 
to indicate 
that \acrshort{CDCVM} was performed.
It is modified in the GET PROCESSING OPTIONS response as well as in the READ RECORD response.
This results in a disagreement~(P1) between the card and terminal on the \TTQx{} and \CTQx{}.

Mastercard transit transactions 
are protected by 
an additional check 
that the \MCCfull{} indicates a transport service.  
The \MCCx{} is authenticated by the \SDADx{}.
Thus, for Mastercard, this attack is limited to relaying
to a transit terminal.
The Visa kernel does not send the \MCCx{} to the card but sends it 
to the acquirer who forwards it to the payment network.
Apple proposed to prevent the attack on Visa's side by checking 
the \MCCx{} for
Apple Pay transactions without \acrshort{CDCVM} (indicated by the \IADx{}, see \secref{SubSec:Protocol:Phone}),

Experiments showed that Samsung Pay
always allows for transit mode transactions.
However these are only allowed for a value of zero, which 
makes the attack useless.

\subsection{Merchant-Holding-The-Bag}\label{SubSec:Attacks:MerchantHoldingBag}
Offline authorized transactions are sent to the issuer after the terminal has accepted the transaction and the cardholder has likely left the shop. 
\cite{basinEMVStandardBreak2021a} report an attack on 
such transactions 
where the attacker can
use their card at a terminal that accepts the transaction and walk out with the goods. 
However, the bank declines the transaction later and the attacker is not charged for the goods. 
Thus, the merchant is left ``holding the bag.''
This violates the no delayed decline property~(P2).

This attack 
targets
low-value
transactions with the \acrshort{ODA} method \acrshort{SDA} and \acrshort{DDA}.
The MAC \ACx{} is modified, which is not authenticated by \acrshort{SDA} and \acrshort{DDA}, see \secref{SubSec:Flaw:ACNotAuth},
and the terminal cannot verify the \ACx{} because it does not have access to the symmetric key~\mkx{}. 
This leads to a disagreement~(P1) between the card and terminal on the \ACx{}.
After the terminal authorized the transaction offline and the attacker walked out with the goods, the bank detects the 
invalid
\ACx{} and declines the transaction.
Note that this attack is not possible for transactions with \acrshort{CDA} 
as \acrshort{CDA} authenticates the \ACx{}.

\subsection{Downgrade Attacks}
A downgrade attack tricks the protocol participants into executing a vulnerable version of the protocol instead of a more secure version that they both support. 
This enables an adversary to perform an attack on the vulnerable version.

\subsubsection{Maestro to Mastercard}\label{SubSec:Attacks:Downgrade:MaestroToMastercard}
PIN-Bypass attack by~\cite{basinInducingAuthenticationFailures2023}, see \secref{SubSection:Attacks:InducingAuthenticationFailure}, could not 
target
Maestro cards directly.
Maestro cards run the Mastercard kernel but with different configuration data than Mastercard cards.
The attack could however be combined with an attack similar to the \emph{Card Brand Mixup} attack by~\cite{basinCardBrandMixup2021}, see~\secref{SubSec:Attacks:PINBypassMastercard:CardBrandMixup}.
The unprotected \AIDxpl{}, see \secref{SubSec:Flaws:AIDNotProtected}, 
are modified so that the terminal runs the Mastercard flow while the card runs the Maestro flow, 
leading to a disagreement~(P1) 
on the \AIDx{}.

\section{Discussion}\label{Section:Discussion}

We discuss here some of our findings regarding the flaws discovered and the methods used to find attacks.

\subsection{Flaw Categories and Causes}\label{Section:Discussion:Flaws}

We now highlight several observations we made while categorizing these flaws and the resulting attacks.

\subsubsection{Authentication}
\looseness=-1
Although the flaws listed in \secref{Section:Flaws} vary considerably, 
missing data authentication
is particularly prominent.
Attacks become possible when 
even 
just one data field is not 
authenticated,
such as the \CTQx{} in \secref{SubSec:Flaws:CTQnotProtected}, the \TTQx{} in \secref{SubSec:Flaws:TTQnotProtected}, or the \AIDx{} in \secref{SubSec:Flaws:AIDNotProtected}.

EMV provides two means to authenticate data: the signature-based \acrshort{ODA} checked by the terminal and the MAC-based \acrshort{OTA} checked by the issuer.
EMV inconsistently handles which data is authenticated by which mechanism.
For example, for Visa transactions, 
\acrshort{OTA} should protect the \TTQx{}, but it does not protect the \CTQx{}.
In contrast, \acrshort{ODA} protects the \CTQx{}, but it does not protect the \TTQx{}.

In addition to 
the differences in signature and MAC authentication,
the signature based ODA methods themselves differ in the 
data they authenticate. 
EMV produces different levels of guarantees when using \acrshort{SDA}, \acrshort{DDA}, and \acrshort{CDA}.
Moreover, the most common \acrshort{ODA} methods of the different kernels vary, namely \acrshort{CDA} for Mastercard and \acrshort{fDDA} for Visa, and do not authenticate the same data.

\subsubsection{Specification Ambiguities}
Multiple flaws stem not from mistakes in the 
specification but rather they arise from optional configuration options or aspects outside of the public specification, such as the issuer's behavior.
For example, incrementing the \ATCx{} should prevent out-of-order transactions.
However, 
not all issuers check 
the \ATCx{}, see \secref{SubSec:Flaws:ATCoutOfOrder}.
Another example is that some terminals set the `CDA Failed' bit in the \TACDenialx{} to zero, see \secref{SubSec:Flaws:TACDenail}, 
which results in 
fraudulent transactions being accepted.

\subsubsection{Sources of Flaws}
Flaws in the EMV protocol can be critical and enable criminals to steal money from banks, cardholders, and merchants.
So why have so many flaws in the EMV protocol been discovered?
And why has the situation seemingly not improved over time?
Here we list potential reasons for this.

First, the requirements on the payment system are not well defined and differ between stakeholders, see~\secref{Section:SecurityProperties}.
A prime example of this is that each payment network has its own kernel and the many configuration options.
Moreover, stakeholders must find a trade-off between different requirements, such as keeping the transaction success rate high while preventing malicious transactions, see \secref{SubSec:Properties:StakeholderRequirements}.
For example, setting the \TACDenialx's `CDA Failed' bit to one might prevent fraudulent transactions but it might also reject valid transactions.

Second, as with many protocols, EMV has grown organically over time.
EMV contactless is based on EMV contact.
Supporting backward compatibility for EMV contact within the new context of \acrshort{NFC}-based contactless transactions also introduced new problems, such as the availability of offline PIN verification, see \secref{SubSec:Flaws:OfflinePINoverNFC}.
Other problems arise as modern payment cards support EMV contactless and still often support even the vulnerable magnetic stripe, see \secref{SubSec:Flaws:MagStripeStillSupported}.

The third and final reason 
is the high complexity of the EMV contactless protocol.
This complexity arises from the protocol's historical development and the numerous kernels and configuration options. 
This complexity leads to both design and implementation issues and it complicates testing and other 
measures that could help 
eliminate flaws.

\subsection{Attack Discovery and Prevention}

This SoK investigates flaws and attacks on EMV contactless that were discovered and publicized.
These flaws and attacks were 
often published with suggested fixes to prevent the particular attacks.
These fixes are mostly lightweight and intended to be implemented with minimal protocol modifications.

\subsubsection{Attack Discovery}\label{SubSec:Discussion:AttackDiscovery}

As observed before, identifying attacks on EMV contactless is challenging given its high complexity.
We describe some of the methods that were used to discover the attacks described in \secref{Section:AttacksOnEMV}.
We specifically highlight two general approaches to attack discovery: testing and model checking.

Testing attacks on live systems raises ethical questions.
The \emph{Merchant-Holding-The-Bag} attack by~\cite{basinEMVStandardBreak2021a}, described in \secref{SubSec:Attacks:MerchantHoldingBag}, was not tested as it would defraud the merchant.
In contrast, PIN-bypass attacks~\cite{basinEMVStandardBreak2021a,basinCardBrandMixup2021,basinInducingAuthenticationFailures2023,raduPracticalEMVRelay2022a} were tested using the researchers' own cards to avoid defrauding others.
However,~\cite{basinInducingAuthenticationFailures2023} point out that, even in these experiments where no one is defrauded, legal questions may arise.
They experienced that their accounts at terminal providers were locked and that their cards would be rejected by certain types of terminals.

To avoid ethical and legal issues, some researchers perform experiments on their own terminal implementations~\cite{emmsDangersVerifyPIN2012}.
For example, \cite{emmsHarvestingHighValue2014} find attacks by testing their own implementation, rather than testing on real cards in the wild.
This eliminates potential ethical issues, but leaves open the question of the effectiveness of the attacks 
in the wild.

In general, testing live systems and testing one's own implementation suffice to discover attacks. 
However, these methods cannot prove that a property holds for a protocol, meaning that a protocol meets some well-defined specification.
More formal approaches can not only identify attacks but also verify protocol properties with respect to a given protocol model.
\cite{basinEMVStandardBreak2021a} created a comprehensive formal model using the state-of-the-art verification tool Tamarin~\cite{schmidtAutomatedAnalysisDiffieHellman2012, meierTAMARINProverSymbolic2013}.
Tamarin can, given a protocol model, desired security properties, and an adversary model, provide formal proofs for those properties that hold as well as find attacks on those properties that do not hold.
The analysis by~\cite{basinEMVStandardBreak2021a} revealed insecure configurations and corresponding attacks.
Moreover, they also produced machine-checked proofs for alternative secure configurations. 
Tamarin can also be used to verify proposed fixes~\cite{basinEMVStandardBreak2021a,basinCardBrandMixup2021,raduPracticalEMVRelay2022a,basinInducingAuthenticationFailures2023}.

\looseness=-1
Note that formal method tools such as model checkers only prove properties of an abstract protocol model.
The model might not cover all possible configurations and protocol details and hence they might not capture all possible attacks.
For example, the models by~\cite{basinEMVStandardBreak2021a} could not capture the \emph{Card Brand Mixup} attack by~\cite{basinCardBrandMixup2021} as they did not account for transactions performed by the VISA kernel being forwarded to the Mastercard network.
Later, the original model was extended to capture such forwarding and to verify the proposed fixes.
Similar extensions were made 
for
other attacks~\cite{raduPracticalEMVRelay2022a, basinInducingAuthenticationFailures2023}.
In contrast, some vulnerable configurations might be part of a model but there might not be any real-world card or terminal with this configuration.
Thus, it is crucial to validate attacks with experiments on actual systems.

Summing up, the combination of both formal verification tools as well as testing live systems is needed to discover attacks, verify fixes, and prove security properties.

\subsubsection{Cat and Mouse Game}

To prevent a given attack, the flaws that enable the attack do not necessarily need to be fixed.
The attack could also be prevented by introducing new measures.
For example, the \emph{Card Brand Mixup} attack~\cite{basinCardBrandMixup2021}, see \secref{SubSec:Attacks:PINBypassMastercard:CardBrandMixup}, exploits the unprotected \AIDx{}, see \secref{SubSec:Flaws:AIDNotProtected}.
Mastercard implemented checks on their network to prevent this attack. 
However, the unprotected \AIDx{} remained and it was subsequently exploited by the \emph{Maestro to Mastercard} downgrade attack~\cite{basinInducingAuthenticationFailures2023}, see \secref{SubSec:Attacks:Downgrade:MaestroToMastercard}.

We see an unfortunately all too familiar escalating battle between the discovery of new attacks and the incorporation of fixes.
Some flaws seem to be exploited repeatedly and fundamental problems such as the lack of authentication and verification of important data seem to remain in the protocol.

\subsubsection{Vision}
Is there any escape from the seemingly never ending spiral of penetrate and patch?
This problem seems to result from, or at least to be magnified by, the tremendous complexity and the need for backwards compatibility of EMV contactless.
Reducing complexity by designing a new protocol without backwards compatibility appears unrealistic for a protocol that is as widely deployed and that has as many different stakeholder requirements as EMV does.
Thus, EMV contactless's complexity will likely remain. 

We believe that the security of such a complex protocol can only be guaranteed using formal methods.
At minimum, verification of existing protocols should be carried out.
In the ideal world, security critical protocols such as EMV should be built, hand-in-hand, with formal models and proofs of their correctness. 
Different methods exist to achieve this.
Some methods analyze protocol models, such as symbolic verification tools like Tamarin, and ProVerif~\cite{blanchetEfficientCryptographicProtocol2001}, while others aim to create verified code~\cite{sprengerRefiningSecurityProtocols2018a, bhargavanDYModularSymbolic2021, ProjectEverest}.
Experience building and verifying models of substantial protocols like EMV\cite{basinEMVStandardBreak2021a, basinCardBrandMixup2021, basinGettingChipCard2025}, TLS~1.3~\cite{cremersComprehensiveSymbolicAnalysis2017a}, 5G AKA \cite{basinFormalAnalysis5G2018} and verification down to code of the TLS 1.3 Record Layer \cite{delignat-lavaudImplementingProvingTLS2017} suggest that existing techniques are up to this 
task.

\section{Conclusion}\label{Section:Conclusion}

In this SoK, we explored attacks on EMV contactless, categorizing them and identifying the exploited flaws.
We also 
provided a comprehensive framework consisting of 
the adversary models and security properties, 
and analyzed attack discovery. 
Although our focus was on the contactless protocol, it could be extended to other EMV technologies, like contact, payment tokenization, or 3D secure, or to other technologies such as non-EMV chip cards.

Research on modern payment cards will continue.
The large number of flaws found in the past suggests that many more flaws and attacks will be found in the future.
The new kernel~8 and future technologies require a thorough analysis, some of which is underway~\cite{basinGettingChipCard2025}.
We hope that our framework of adversary models and security properties will be helpful in this regard.

\bibliographystyle{plain}
\bibliography{SoK_final}

\appendix

\section{Message Sequence Charts}\label{Appending:MSC}\label{Appendix:MSC}
\begin{figure*}[tbh]
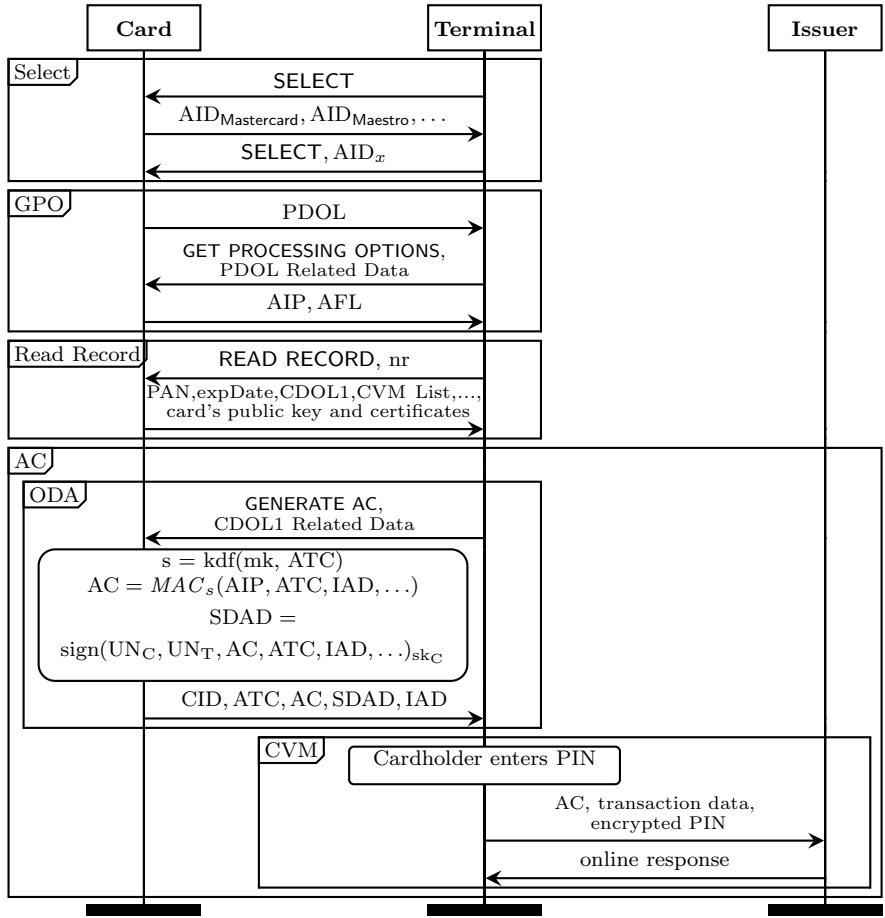

    \centering
    \include{mastercard}
    \caption{Overview of a Mastercard contactless transaction using the \acrshort{ODA} method \acrshort{CDA} and the \acrshort{CVM} method Online PIN.}
    \label{figure:Protocol:Mastercard}
\end{figure*}

See Figures~\ref{figure:Protocol:Mastercard} and \ref{figure:Protocol:Visa} for an overview of kernels 2 and 3.

\begin{figure*}[tbh]
    \centering
    \include{visa}
    \caption{Overview of a Visa contactless transaction using its \acrshort{ODA} method \acrshort{fDDA} and the \acrshort{CVM} method Online PIN.}
    \label{figure:Protocol:Visa}
\end{figure*}

\section{Mag-Stripe Mode}\label{Appendix:MagStripeMode}
EMV contactless offers two modes.
We previously covered EMV contactless's EMV mode. 
Here we provide an overview of EMV contactless's mag-stripe mode, its flaws and attacks on it.
Note that mag-stripe mode is a mode of EMV contactless and operates over NFC.
It is not to be confused with the physical magnetic stripe found on most payment cards.

\subsection{Background Mag-Stripe Mode}\label{SubSec:Protocol:MagStripeMode}
Mag-stripe mode is based on the data on the magnetic stripe found on most payment cards.
During a mag-stripe mode transaction, the card provides the static data that is also stored on the card's magnetic stripe and an additional dynamic verification code over NFC.
Mag-stripe mode is a legacy mode. 
Visa removed mag-stripe mode from their specification in 2016 and it is not part of the new kernel 8.
However, the Mastercard specification still contains mag-stripe mode~\cite{emvcoEMVContactlessSpecifications2023i}.

A mag-stripe mode transaction starts, like EMV mode, with the SELECT and GET PROCESSING OPTIONS command and response.
The card indicates with the \AIPx{} whether it supports EMV mode.
If the card and terminal both support EMV mode, it is used. 
Otherwise, they perform mag-stripe mode.

Similarly to the magnetic stripe, in mag-stripe mode, the card's data is sent as the \TrackOneTwodatax{}. 
This data contains the PAN, expiry date, service code, and so-called discretionary data. 
The \TrackOneTwodatax{} is accessed via the READ RECORD command.
Note that the EMV mode supports the so-called \TrackTwoeqdatax{}, which is similar to the \TrackTwodatax{}. 

In mag-stripe mode transactions, the card authenticates using the \CVCthreefull{}.
However, the \CVCthreex{} does not authenticate the transaction details.
To prevent replay attacks, the \CVCthreex{} is derived from the counter \ATCx{} and the terminal's nonce \UNTx{}.

\subsection{Flaws of Mag-Stripe Mode}\label{SubSection:Flaws:MagStripeMode}

We next list the flaws discovered in mag-stripe mode.
\subsubsection{Weak Random Number Generation}\label{SubSec:Flaws:UNinMagStripeMode} \label{SubSec:Flaws:WeakRand}

Mag-stripe mode's \CVCthreex{} cryptogram is computed over the terminal-sourced nonce \UNTx{}.
\cite{rolandCloningCreditCards2013} discovered a design flaw that reduces the \UNTx{}'s entropy to only 3 decimal digits.
The four-byte \UNTx{} is limited by two factors: the \acrfull{BCD} encoding and the size limitation provided by values stored in the card's mag-stripe data file.
\cite{rolandCloningCreditCards2013} exploit this weakness in their cloning attacks, see \secref{SubSec:Attacks:MagStripeModeClone}.

\subsubsection{AIP Not Protected}\label{SubSec:Flaws:AIPNotProtectedMagStripeMode}

The terminal chooses EMV or mag-stripe mode based on the card sourced \AIPx{}.
However, the \AIPx{} is not protected in mag-stripe mode and can thus be modified by the adversary.
The downgrade attack by~\cite{rolandCloningCreditCards2013} exploits this, see \secref{SubSec:Attacks:MagStripeModeDowngrade}.

\subsubsection{Terminals Retrying Different UNs}\label{SubSec:Flaws:MagStripe:UNRetry}

\cite{rolandCloningCreditCards2013} discovered that in mag-stripe mode some terminals retry transactions with a different nonce \UNTx{} if the \CVCthreex{} cryptogram verification fails. 
They showed 
that some terminals have this behavior, while others do not.
This flaw allows for an optimization of the 
cloning attack by~\cite{rolandCloningCreditCards2013}, see \secref{SubSec:Attacks:MagStripeModeClone}.

\subsection{Attacks on Mag-Stripe Mode}\label{SubSec:Attacks:MagStripeMode}

We next list the attacks on mag-stripe mode.
As stated in the initial paragraph of~\secref{Appendix:MagStripeMode}, mag-stripe mode is not to be confused with the magnetic stripe.

\subsubsection{Eavesdropping on Card Data}\label{SubSec:Attacks:EavesdroppCardData}

The attack by~\cite{emmsPracticalAttackContactless2011} eavesdrops on the transaction between a terminal and a card via a hidden \acrshort{NFC} reader placed at a Point-of-Sales terminal.
The attacker can record the card number, cardholder name, expiry date, and issue date, as the card sends this data unencrypted, see \secref{SubSec:Flaws:UnencryptedData}. 
This data can be used for card-not-present transactions, such as payments over a website or the phone. 
This attack violates the secrecy of this data~(P5).
As it does not forge an EMV contactless transaction, no further properties are violated.

As~\cite{emmsPracticalAttackContactless2011} are not clear on the EMV mode, we presume
that this data was retrieved using mag-stripe mode.
This presumption is based on the kernel~2 and~3 specifications~v2.1 from 2011~\cite{emvcoEMVContactlessSpecifications2011e}, which is the earliest version available on the EMVCo website.

Online purchases often require the \CSCx{}, which is not part of a EMV contactless transaction, but is printed on the card's backside. 
\cite{emmsDangersVerifyPIN2012} state that card-not-present transactions can even be performed in many cases without the \CSCx{}.
Otherwise they propose to capture the \CSCx{} using a hidden camera~(A8).

\subsubsection{Mag-Stripe Mode Cloning}\label{SubSec:Attacks:MagStripeModeClone}

The \emph{Mag-Stripe Mode Cloning} attack~\cite{rolandCloningCreditCards2013} uses skimmed data from a Mastercard mag-stripe mode transaction to create a functional clone supporting mag-stripe mode. 
The \CVCthreex{} is computed over the card-sourced counter \ATCx{} and the terminal-sourced nonce \UNTx{} to prevent replays.
However, due to a design flaw, \UNTx{}'s entropy  is only 3 decimal digits, see \secref{SubSec:Flaws:UNinMagStripeMode}.
The experiments of~\cite{rolandCloningCreditCards2013} showed that a card could be queried for those 1000 nonces in about one minute. 
The recorded \CVCthreexpl{} can then be stored on the card clone.
This attack violates the secrecy~(P5) of the mag-stripe mode data and as the card clone can be used in EMV contactless transactions, this also violates the cardholder intent~(P3). 

The attack can be further improved since some terminals retry transactions with a different \UNTx{} if the \CVCthreex{} verification fails, see \secref{SubSec:Flaws:MagStripe:UNRetry}.
This allows the adversary to perform the attack without querying the \CVCthreex{} for every possible \UNTx{}.
The cloned card can send an error message after receiving a \UNTx{} that it did not record and it will receive a different \UNTx{} from the terminal.

As stated above, the \CVCthreex{} cryptogram is also calculated over the counter \ATCx{}.
The authors of~\cite{rolandCloningCreditCards2013} assume that issuers record the \ATCxpl{} used for each card to reject out-of-order \ATCxpl{}. 
This would limit the attacker's window of opportunity. 
However,~\cite{gallowayFirstContactNew2019} showed that the \ATCx{} is not properly checked by many issuing banks, see \secref{SubSec:Flaws:ATCoutOfOrder}.

\cite{rolandCloningCreditCards2013} implemented a proof-of-concept system, consisting of an Android app to record the \CVCthreex{} and a Java Card application for card cloning.

\subsubsection{EMV Mode to Mag-Stripe Mode}\label{SubSec:Attacks:MagStripeModeDowngrade}

The \emph{Mag-Stripe Mode Cloning} attack~\cite{rolandCloningCreditCards2013}, see \secref{SubSec:Attacks:MagStripeModeClone},
relies on the card and terminal choosing mag-stripe mode. 
However, if the card and terminal both support EMV mode, they would choose EMV mode over mag-stripe mode. 
The authors extend their attack by tricking the terminal into using mag-stripe mode with a card clone, even though the original card supports EMV mode. 

The \AIPx{} 
is modified to indicate that no EMV mode is supported. 
This modification is possible as mag-stripe mode does not authenticate the \AIPx{}, see \secref{SubSec:Flaws:AIPNotProtectedMagStripeMode}.
This leads to a disagreement on the \AIPx{}~(P1).
In combination with the \emph{Mag-Stripe Mode Cloning} attack, an attacker can use the cloned card at all terminals that support mag-stripe mode, 
even though the original card supports EMV mode.

\section{Tables}\label{Appendix:Tables}
Table \ref{table:allAttacks} summarizes the attacks presented in this SoK.

\begin{landscape}

    \begin{table}[]
        \centering
        \resizebox{\columnwidth}{!}{
        \begin{tabular}{|P{4.1cm}|P{9.3cm}|P{1.604cm}|P{9.225cm}|P{1.3cm}|P{1.2cm}|}
        \hline
        Attack&Violated Security Properties&Adv. Cap.&Flaws&Demon- strated&Patched\\ \hline
        \ref{SubSec:Attacks:MagStripeClone} Magnetic Stripe Cloning \cite{gallowayItOnlyTakes2020} &   P5) Secrecy (Track 1, 2 Eqv. Data) &  A1) NFC  & \ref{SubSec:Flaws:MagStripeStillSupported} Mag. Stripes Are Still Supp., \ref{SubSec:Flaws:UnencryptedData} Unencr. Data, \ref{SubSec:Flaws:MagStripeDataInEMV} Mag. Stripe Data in EMV, \ref{SubSec:Flaws:SignatureInterface} Weakness of Paper Sign.& in the wild  & \\ \hline
        \ref{SubSec:Attacks:ForeignCurrency} No Cardholder Verif. in Foreign Currencies \cite{emmsHarvestingHighValue2014} &  P3) Cardh. Int., P3.1) High-Value Trans. Require CVM, P3.2) Card Close to Term. &  A1) NFC, A5) Term.
         & \ref{SubSec:Flaws:ForeignCurrency} No Limit for Foreign Currencies, \ref{SubSec:Flaws:MerchantNotAuth} Merchant Details Are Not Auth. & in exp. setting & \\ \hline
        \ref{SubSec:Attacks:ReplayUNReuse} Replay with Nonce Reuse \cite{gallowayFirstContactNew2019} &   P1) Data Auth. (replay) P3) Cardh. Int., P3.2) Card Close to Term. & A1) NFC, A5) Term.
         & \ref{SubSec:Flaws:ATCoutOfOrder} Out-Of-Order ATCs Accepted, \ref{SubSec:Flaws:UNReuse} UN Reuse Not Prevented & in the wild & \\ \hline
        \ref{SubSec:Attacks:PINGuess} PIN Guessing \cite{emmsDangersVerifyPIN2012,emmsRisksOfflineVerify2013} &  P5) Secrecy (PIN), P3) Cardh. Int. &  A1) NFC  & \ref{SubSec:Flaws:OfflinePINoverNFC} PIN Veriﬁcation Over NFC& in exp. setting &  \\ \hline
        \ref{Subsec:Attacks:PINSpamming} PIN-Guess-Spam. \cite{emmsDangersVerifyPIN2012,emmsRisksOfflineVerify2013} & P6) Availability & A1) NFC & \ref{SubSec:Flaws:OfflinePINoverNFC} PIN Veriﬁcation Over NFC&  in exp. setting & \\ \hline
        \ref{SUBSEC:Attacks:FirstVisaPINBypass} Combined TTQ, CTQ Mod. Attack \cite{gallowayFirstContactNew2019} &   P1) Data Auth. (TTQ, CTQ), P3) Cardh. Int., P3.1) High- Value Trans. Require CVM, P3.2) Card Close to Term. &  A1) NFC  & \ref{SubSec:Flaws:CTQnotProtected} CTQ is Not Prot., \ref{SubSec:Flaws:TTQnotProtected} TTQ is Not Prot. \ref{SubSec:Falaws:PlasticCardsCDCVM} Not Checked if Plastic Cards Do \acrshort{CDCVM} \ref{SubSec:Flaws:Relay} No Relay Prot. & in the wild  & No \cite{basinInducingAuthenticationFailures2023} \\ \hline
        \ref{SubSec:Attacks:2ndVisaPINBypass} CTQ Mod. Attack  \cite{basinEMVStandardBreak2021a} & P1) Data Auth. (CTQ), P3) Cardh. Int., P3.1) High- Value Trans. Require CVM, P3.2) Card Close to Term. & A1) NFC  &  \ref{SubSec:Flaws:CTQnotProtected} CTQ is Not Prot., \ref{SubSec:Falaws:PlasticCardsCDCVM} No Check if Plastic Cards do CDCVM, \ref{SubSec:Flaws:Relay} No Relay Prot. & in the wild  & No \cite{basinInducingAuthenticationFailures2023} \\ \hline
        \ref{SubSec:Attacks:PINBypassMastercard:CardBrandMixup} Card Brand Mixup \cite{basinCardBrandMixup2021} &   P1) Data Auth. (AID, CTQ), P3) Cardh. Int., P3.1) High-Value Trans. Require CVM, P3.2) Card Close to Term. &  A1) NFC  &  \ref{SubSec:Flaws:VisaFromMastercard} Visa Resp. Built from Mastercard Resp., \ref{SubSec:Falaws:PlasticCardsCDCVM} No Check if Plastic Cards Do \acrshort{CDCVM}, \ref{SubSec:Flaws:CheckAIDPAN} No Check if AID and PAN Match, \ref{SubSec:Flaws:AIDNotProtected} AID is Not Prot., \ref{SubSec:Flaws:CTQnotProtected} CTQ is Not Prot., \ref{SubSec:Flaws:Relay} No Relay Prot. & in the wild & Yes~\cite{basinCardBrandMixup2021} \\ \hline
        \ref{SubSection:Attacks:InducingAuthenticationFailure} Inducing Auth. Failure \cite{basinInducingAuthenticationFailures2023} & P1) Data Auth. (CA PK Index, CVM List, IAC-Denial), P3) Cardh. Int., P3.1) High-Value Trans. Require CVM, P3.2) Card Close to Term. &  A1) NFC  & \ref{SubSec:Flaws:TACDenail} TAC-Denial Set to Zero, \ref{SubSec:Flaws:SignatureInterface} Weakness of Paper Sign., \ref{SuSec:Flaws:CALookupFailure} CA Look up Failure Not Declined, \ref{SubSec:Flaws:Relay} No Relay Prot. & in the wild & No~\cite{basinInducingAuthenticationFailures2023}\\ \hline
        \ref{SubSec:Attacks:FirstVisaPINBypassGooglePay} TTQ, CTQ Mod. Attack on Google Pay \cite{gallowayFirstContactNew2019} & P1) Data Auth.(TTQ), P3) Cardh. Int., P3.1) High-Value Trans. Require CVM, P3.2) Card Close to Term. &  A1) NFC  & \ref{SubSec:Flaws:TTQnotProtected} TTQ is Not Prot., \ref{SubSub:Flaws:SendingCDCVM} Phones Always Send CDCVM \ref{SubSec:Flaws:Relay} No Relay Prot. & in the wild & No \cite{raduPracticalEMVRelay2022a}\\ \hline
        \ref{SubSec:Attacks:MagicByte} “Magic byte” PIN-Bypass \cite{raduPracticalEMVRelay2022a} & P1) Data Auth. (TTQ, CTQ), P3) Cardh. Int., P3.1) High-Value Trans. Require CVM, P3.2) Card Close to Term. &  A1) NFC  & \ref{SubSec:Flaws:CTQnotProtected} CTQ is Not Prot., \ref{SubSec:Flaws:TTQnotProtected} TTQ is Not Prot., \ref{SubSec:Flaws:MagicByte} Transit Mode Init. with “Magic Byte” \ref{SubSec:Flaws:Relay} No Relay Prot. & in the wild  & No \cite{raduPracticalEMVRelay2022a}\\ \hline
        \ref{SubSec:Attacks:MerchantHoldingBag} Merchant-Holding-The-Bag \cite{basinEMVStandardBreak2021a} & P1) Data Auth. (AC), P2) No Delayed Decline, P3.2) Card Close to Term. &  A1) NFC  & \ref{SubSec:Flaw:ACNotAuth} SDA and DDA do Not Auth. the AC, \ref{SubSec:Flaws:Relay} No Relay Prot.  & No & \\ \hline
        \ref{SubSec:Attacks:Downgrade:MaestroToMastercard} Maestro to Mastercard \cite{basinInducingAuthenticationFailures2023} &   P1) Data Auth. (AID), P3.2) Card Close to Term. &  A1) NFC  & \ref{SubSec:Flaws:AIDNotProtected} AID is Not Prot., \ref{SubSec:Flaws:Relay} No Relay Prot. & in the wild & No \cite{basinInducingAuthenticationFailures2023}\\ \hline
        \ref{SubSec:Attacks:EavesdroppCardData} Eavesdropping on Card Data \cite{emmsPracticalAttackContactless2011} & P5) Secrecy (data for card not present transactions)  & A1) NFC, A8)Visual & \ref{SubSec:Flaws:UnencryptedData} Unencr. Data& in the wild &  mag-stripe \\ \hline
        \ref{SubSec:Attacks:MagStripeModeClone} Mag-Stripe Mode Cloning \cite{rolandCloningCreditCards2013} & P5) Secrecy (mag-stripe mode data), P3) Cardh. Int., P3.2) Card Close to Term. & A1) NFC  & \ref{SubSec:Flaws:ATCoutOfOrder} Out-Of-Order ATCs Accepted, \ref{SubSec:Flaws:WeakRand} Weak Random Number Generation, \ref{SubSec:Flaws:MagStripe:UNRetry} Terminals Retrying Different UNs & in the wild &  mag-stripe \\ \hline
        \ref{SubSec:Attacks:MagStripeModeDowngrade} EMV Mode to Mag-Stripe Mode\cite{rolandCloningCreditCards2013} &   P1) Data Auth. (AIP), P3.2) Card Close to Term. &  A1) NFC  & \ref{SubSec:Flaws:AIPNotProtectedMagStripeMode} AIP Not Prot. & in the wild & mag-stripe \\ \hline
        \end{tabular}
        }
        \caption{Overview of the attacks covered. For each attack, we list the security properties violated, the adversary capabilities required to perform the attack, the flaws that the attack exploits,
        if the attack was demonstrated in the wild or in an experimental setting and if measures preventing the attack were implemented.   
        The security properties refer to the properties P1) - P6) listed in~\secref{SubSec:SecProp:SystProp}. Adversary capabilities A1), A5), A8) refer to~\secref{SubSec:Adv:Cap}.
        The column \emph{patched} contains the citation of the latest information available. 
        Note that the attack could have been patched in the meantime.    
        The attacks labeled with \emph{mag-stripe} rely on the legacy mag-stripe mode.
        The \emph{patched} field is left empty for the attacks for which we do not have any information.
        We use short-hands such as \emph{Cardh. Int.} for Cardholder Intent, \emph{Auth.} for Authentication or authenticate, \emph{Trans.} for Transaction, \emph{Sign.} for Signature, \emph{prot.} for protected or protection, \emph{Unencr.} for Unencrypted, and \emph{Term.} for Terminal. 
           \label{table:allAttacks}}
        
\end{table}

\end{landscape}

\end{document}